\documentclass{article}
\usepackage{graphicx}
\usepackage{epstopdf, epsfig}
\usepackage{xcolor}

\usepackage{import} %

\newcommand{\vhf}{\bf{\hat {f}}}
\newcommand{\vhg}{\bf{\hat {g}}}

\newcommand{\vhhk}{\bf{{\hat{h}}^0_{\vk}}}
\newcommand{\vhhp}{\bf{{\hat{h}}^+_{\vk}}}
\newcommand{\vhhm}{\bf{{\hat{h}}^-_{\vk}}}
\newcommand{\vhhpm}{\bf{{\hat{h}}^\pm_{\vk}}}
\newcommand{\vhhs}[2]{\bf{{{\hat{h}}}^{#1}_{#2}}}

\newcommand{\vhe}{{\bf{\hat{e}}}}

\newcommand{\dvol}{dV}

\newcommand{\beqa}{\begin{eqnarray}}
\newcommand{\eeqa}{\end{eqnarray}}
\newcommand{\beq}{\begin{equation}}
\newcommand{\eeq}{\end{equation}}

\newcommand{\beqal}{\begin{equation}\begin{aligned}}
\newcommand{\eeqal}{\end{aligned}\end{equation}}

\newcommand{\pat}{\partial_t}

\newcommand{\vf}{\bf{{f}}}

\newcommand{\vg}{\bf{{g}}}

\newcommand{\vk}{\bf{k}}
\newcommand{\vm}{\bf{m}}

\newcommand{\vp}{\bf{p}}
\newcommand{\vq}{\bf{q}}

\newcommand{\vx}{\bf{x}}

\newcommand{\vlambda}{{\boldsymbol \lambda}}
\newcommand{\vmu}{{\boldsymbol \mu}}

\newcommand{\bit}{\begin{itemize}}
\newcommand{\eit}{\end{itemize}}

\newcommand{\neutrindex}{m}

\newcommand{\neutrfield}{\vf}
\newcommand{\neutrfieldB}{\vg}
\newcommand{\neutrfieldF}{\vhf}
\newcommand{\neutrfieldBF}{\vhg}

\newcommand{\vneutrindex}{{\bf \neutrindex}}

\newcommand{\neutrenergysymb}{{\cal{E}}}

\newcommand{\rmsM}{{\cal{M}}} 
\newcommand{\kspec}{K} 

\newcommand{\hel}{{\cal{H}}}
\newcommand{\mhel}{{\hel^M}} 
\newcommand{\mhelK}{{\hel^M_K}} 

\newcommand{\mhelF}{H^M} 
\newcommand{\mhelFK}{\mhelF_\kspec} 
\newcommand{\mhelFk}{\mhelF_{\vk}} 

\newcommand{\emagF}{E^M} 
\newcommand{\emagFhSKprep}{E^{M,\helsignK}} 
\newcommand{\emagFhSQprep}{E^{M,\helsignQ}} 

\newcommand{\emagFhSKZ}{\emagFhSKprep_{\KZ}} 
\newcommand{\emagFhSQ}{\emagFhSQprep_Q} 

\newcommand{\emagFK}{\emagF_\kspec} 

\newcommand{\sekinF}{E^V} 
\newcommand{\sekinFhS}{E^{V,\helsignP}} 
\newcommand{\sekinFhP}{E^{V,P}} 
\newcommand{\sekinFhC}{E^{V,C}} 
\newcommand{\sekinFK}{\sekinF_\kspec} 

\newcommand{\sekinFhSP}{\sekinFhS_P} 

\newcommand{\ttt}{t_{\cal{T}}} 
\newcommand{\Lbox}{L} 

\newcommand{\Einj}{{\epsilon^{K}_{inj}}} 
\newcommand{\EMinj}{{\epsilon^{M}_{inj}}} 
\newcommand{\mhelinj}{{\epsilon^{\mhel}_{inj}}} 

\newcommand{\emag}{\neutrenergysymb^M} 
\newcommand{\Ekin}{\neutrenergysymb^K} 
\newcommand{\Emag}{\neutrenergysymb^M} 
\newcommand{\ekin}{\Ekin} 

\newcommand{\epot}{\neutrenergysymb^\rho} 

\newcommand{\FmagL}{b} 
\newcommand{\Fmag}{{\bf{\FmagL}}} 
\newcommand{\FmagK}{\Fmag_K} 

\newcommand{\FmagQ}{\Fmag_Q} 
\newcommand{\FmagLF}{{{}\hat{b}}} 
\newcommand{\FmagF}{\bf{\FmagLF}} 
\newcommand{\FmagFL}{\FmagL} 
\newcommand{\FmagAL}{a} 
\newcommand{\FmagA}{{\bf{\FmagAL}}} 
\newcommand{\FmagAK}{\FmagA_K} 

\newcommand{\FvelL}{v} 
\newcommand{\Fvel}{{\bf{\FvelL}}} 
\newcommand{\FvelP}{\Fvel_P} 

\newcommand{\FvelFL}{\FvelL} 

\newcommand{\FvelhC}{\Fvel^C} 
\newcommand{\FvelhP}{\Fvel^P}
\newcommand{\FvelhN}{\Fvel^N}
\newcommand{\FvelhX}{\Fvel^X}
\newcommand{\FvelhXP}{\Fvel^X_P}

\newcommand{\FmaghP}{\Fmag^P}
\newcommand{\FmaghN}{\Fmag^N}

\newcommand{\FmagXX}[2]{\Fmag^{#1}_{#2}}
\newcommand{\FvelXX}[2]{\Fvel^{#1}_{#2}}

\newcommand{\helsign}{s}
\newcommand{\helsignm}{\helsign_m}
\newcommand{\helsignk}{\helsign_k}
\newcommand{\helsignp}{\helsign_p}
\newcommand{\helsignq}{\helsign_q}

\newcommand{\hsm}{\helsignm}
\newcommand{\hsk}{\helsignk}
\newcommand{\hsp}{\helsignp}
\newcommand{\hsq}{\helsignq}

\newcommand{\helsignK}{\helsign_K}
\newcommand{\helsignP}{\helsign_P}
\newcommand{\helsignQ}{\helsign_Q}

\newcommand{\hsK}{\helsignK}
\newcommand{\hsP}{\helsignP}
\newcommand{\hsQ}{\helsignQ}

\newcommand{\FmagFhvkP}{{\FmagLF^{+}_{\vk}}}
\newcommand{\FmagFhvkM}{{\FmagLF^{-}_{\vk}}}
\newcommand{\FmagFhvkPM}{{\FmagLF^{\pm}_{\vk}}}
\newcommand{\FmagFhvkPMvec}{{{\bf{\FmagLF}}^{\pm}_{\vk}}}

\newcommand{\triadgeom}{g}
\newcommand{\triadgeomkpqC}{\triadgeom^{\helsignk=S,0,\helsignq=\pm S}_{k,p,q}}
\newcommand{\triadgeomkpq}{\triadgeom^{\helsignk,\helsignp,\helsignq}_{k,p,q}}

\newcommand{\FvelA}{{\bf{\FvelL_A}}} 

\newcommand{\cs}{c_s} 

\newcommand{\kone}{\kappa} 

\newcommand{\rhowIIv}{\bf{w}} 

\newcommand{\IscHm}{{\cal{I}}_{\mhel}}

\newcommand{\mhelfracFor}{h_f}

\newcommand{\rhoz}{\rho_0}

\newcommand{\mhlkinj}{\kspec^M_f}

\newcommand{\helP}{P}
\newcommand{\helC}{C}
\newcommand{\helN}{N}
\newcommand{\helX}{X}

\newcommand{\helS}{S}
\newcommand{\helSA}{\helS}
\newcommand{\helSB}{\helS}

\newcommand{\PXP}{\helP\helX\helP}

\newcommand{\PPP}{\helP\helP\helP}
\newcommand{\PCP}{\helP\helC\helP}
\newcommand{\PNP}{\helP\helN\helP}
\newcommand{\NPN}{\helN\helP\helN}
\newcommand{\NCN}{\helN\helC\helN}
\newcommand{\NNN}{\helN\helN\helN}
\newcommand{\SPS}{\helSA\helP\helSB}
\newcommand{\SCS}{\helSA\helC\helSB}
\newcommand{\SNS}{\helSA\helN\helSB}

\newcommand{\KZ}{K_0}

\newcommand{\NTsf}{{\cal{U}}}
\newcommand{\MTsf}{{\cal{M}}}

\newcommand{\Tsf}{{\cal{T}}}

\newcommand{\TsfMhl}{\Tsf^{\mhel}}
\newcommand{\NormTsfMhl}{{\cal{N}}^{\mhel}}
\newcommand{\NTsfMhl}{\NTsf^{\mhel}}
\newcommand{\MTsfMhl}{\MTsf^{\mhel}}

\newcommand{\TsfMhlQK}{\TsfMhl(Q,K)}

\newcommand{\TsfMhlQKZv}[1]{\TsfMhl(Q,#1)}
\newcommand{\TsfMhlPK}{\MTsfMhl(P,K)}
\newcommand{\TsfMhlQPK}{\TsfMhl(Q,P,K)}
\newcommand{\TsfMhlHELQPK}{\TsfMhl_{\helsignK\helsignP\helsignQ}(Q,P,K)}

\newcommand{\TsfMhlHELQPKZ}[2]{\TsfMhl_{#1}(Q,P,#2)}
\newcommand{\NormTsfMhlHELQPKZ}[2]{\NormTsfMhl_{#1}(Q,P,#2)}
\newcommand{\TsfMhlHELBASEQPKZ}[1]{\TsfMhlHELQPKZ{\helsignK\helsignP\helsignQ}{#1}}
\newcommand{\TsfMhlHELBASEQPKZnormmodule}[1]{\NormTsfMhlHELQPKZ{\helsignK\helsignP\helsignQ}{#1}}

\newcommand{\TsfMhlHEL}[1]{\Tsf^{\mhel}_{#1}}
\newcommand{\TsfMhlPPP}{\TsfMhlHEL{\PPP}}
\newcommand{\TsfMhlPCP}{\TsfMhlHEL{\PCP}}
\newcommand{\TsfMhlPNP}{\TsfMhlHEL{\PNP}}

\newcommand{\TsfMhlXXX}[1]{\TsfMhlHEL{#1}}

\newcommand{\TsfMhlPPPQK}{\TsfMhlHEL{\PPP}(Q,K)}
\newcommand{\TsfMhlPCPQK}{\TsfMhlHEL{\PCP}(Q,K)}
\newcommand{\TsfMhlPNPQK}{\TsfMhlHEL{\PNP}(Q,K)}

\newcommand{\TsfMhlPPPQKZv}[1]{\TsfMhlHEL{\PPP}(Q,#1)}
\newcommand{\TsfMhlPCPQKZv}[1]{\TsfMhlHEL{\PCP}(Q,#1)}

\newcommand{\BBsuf}{bb}
\newcommand{\VBsuf}{vb}
\newcommand{\KBTsuf}{vbT}
\newcommand{\KBPsuf}{vbP}

\newcommand{\BVsuf}{bv}

\newcommand{\TsfMhlBB}{\NTsfMhl_{\BBsuf}}
\newcommand{\TsfMhlBBQK}{\TsfMhlBB(Q,K)}

\newcommand{\TsfMhlVB}{\NTsfMhl_{\VBsuf}}

\newcommand{\TsfMhlKBT}{\NTsfMhl_{\KBTsuf}}
\newcommand{\TsfMhlKBTQK}{\TsfMhlKBT(Q,K)}

\newcommand{\TsfMhlKBP}{\NTsfMhl_{\KBPsuf}}
\newcommand{\TsfMhlKBPQK}{\TsfMhlKBP(Q,K)}

\newcommand{\TsfEm}{\Tsf^{\emag}}

\newcommand{\TsfEmcomp}{\TsfEm}
\newcommand{\NTsfEm}{\NTsf^{\emag}}
\newcommand{\MTsfEm}{\MTsf^{\emag}}

\newcommand{\TsfEmVBcomp}{\TsfEmcomp_{\VBsuf}}

\newcommand{\TsfEmBBvhX}{\MTsfEm_{\BBsuf,\FvelhX}}
\newcommand{\NTsfEmBB}{\NTsfEm_{\BBsuf}}

\newcommand{\TsfEmBBcomp}{\TsfEmcomp_{\BBsuf}}
\newcommand{\MTsfEmBB}{\MTsfEm_{\BBsuf}}

\newcommand{\TsfEmBBQPK}{\NTsfEmBB(Q,P,K)}
\newcommand{\TsfEmKBT}{\TsfEm_{\KBTsuf}}
\newcommand{\TsfEmKBTvhX}{\TsfEm_{\KBTsuf,\FvelhX}}
\newcommand{\TsfEmKBTQPK}{\NTsfEm_{\KBTsuf}(Q,P,K)}

\newcommand{\TsfEmKBPQPK}{\NTsfEm_{\KBPsuf}(Q,P,K)}
\newcommand{\TsfEmKBTPK}{\TsfEm_{\KBTsuf}(P,K)}

\newcommand{\TsfEmKBP}{\TsfEm_{\KBPsuf}}

\newcommand{\TsfEmKBPvhC}{\TsfEm_{\KBPsuf,\FvelhC}}
\newcommand{\TsfEmKBPPK}{\TsfEm_{\KBPsuf}(P,K)}

\newcommand{\TsfEmBBPK}{\MTsfEmBB(P,K)}

\newcommand{\TsfEk}{\Tsf^{\ekin}}

\newcommand{\TsfEkcomp}{\TsfEk}

\newcommand{\TsfEkBVcomp}{\TsfEkcomp_{\BVsuf}}

\newcommand{\TsfEkBVKPcomp}{\TsfEkBVcomp(K,P)}
\newcommand{\TsfEmBBQKcomp}{\TsfEmBBcomp(Q,K)}
\newcommand{\TsfEmVBPKcomp}{\TsfEmVBcomp(P,K)}

\newcommand{\TsfEmBBQPKcomp}{\TsfEmBBQPK}

\newcommand{\TsfEmBBKQcomp}{\TsfEmBBcomp(K,Q)}

\newcommand{\QgeomL}{G}

\newcommand{\VBa}{V \leftrightarrow B}
\newcommand{\BBa}{B \leftrightarrow B}

\newcommand{\uncurl}{\mathrm{\bf rot}^{-1}}

\newcommand{\sm}{\text{-}}

\newcommand{\ForceV}{\bf{f}_V}

\newcommand{\ForceM}{\bf{f}_M}

\newcommand{\mathI}{{\bf I}}

\newcommand{\volavg}[1]{\langle #1 \rangle}

\newcommand{\lra}{\leftrightarrow}

\newcommand{\MVIIc}{M8c}

	\def\arrlength{4.}
	\def\arrXorig{0.}

	\def\arrXoffsKP{0.2}

	\def\arrTopY{1.}
	\def\arrBotY{0.}

	\def\xMAGss{0.67}
	\def\xMAGis{0.2}

	\def\markYsize{0.1}
	\newcommand{\intermediateNLITA}{(2.95,1.2)}
	\newcommand{\intermediateNLITB}{(2.75,1.2)}

	\def\LITAx{2.3}
	\def\LITBx{2.}
	\def\LITCx{1.7}
	\def\LITDx{1.4}

	\def\LITAy{\arrTopY}
	\def\LITBy{\arrTopY}
	\def\LITCy{\arrTopY}
	\def\LITDy{\arrTopY}

	\newcommand{\deflenARR}{2.}

	\newcommand{\finer}{0.1}
	\newcommand{\normthick}{0.5}
	\newcommand{\thicker}{1.}
	
	\newcommand{\thickBase}{0.5}

	\newcommand{\arrowKBT}{-{Latex[length=\deflenARR]}}
	\newcommand{\arrowKBP}{-{Latex[length=\deflenARR,red]}}
	\newcommand{\colorKBT}{black}
	\newcommand{\colorKBP}{red}

	\newcommand{\txtSizeTsfEmSketch}{\tiny}

	\def\offsXLIT{-0.3}

\title{Inverse transfer of magnetic helicity in direct numerical simulations of compressible isothermal turbulence: helical transfers}

\author{Jean-Mathieu Teissier$^1$ and Wolf-Christian M\"uller$^{1,2}$}

\date{\small $^1$ Technische  Universit\"{a}t Berlin, ER 3-2, Hardenbergstr. 36a, D-10623 Berlin, Germany \\
$^2$ Max-Planck/Princeton Center for Plasma Physics\\ \vspace{1em}
\today}

\newcommand{\upi}{\pi}
\usepackage{amsmath}
\usepackage{amssymb}

\newcommand{\grpath}{figures/}
\graphicspath{{\grpath}}
\definecolor{corrclr}{rgb}{0,0,0}
\definecolor{corrclrT}{rgb}{0,0,0}
\newcommand{\corr}[1]{\textcolor{corrclr}{#1}}
\newcommand{\corrT}[1]{\textcolor{corrclrT}{#1}}

\def\plotvspaceval{0}
\def\plotvspacevalend{0}

\def\vspacetikzfig{0}

\usepackage{tikz}
\usepackage{pgfplots}
\pgfplotsset{width=7cm,compat=1.8}

\usepackage{tkz-euclide}
\usetikzlibrary{arrows.meta}

\newcommand{\spectransMainInstant}{\IscHm \approx \frac{\Lbox}{10}}

\newcommand{\tsdefault}{\Large}

\newcommand{\tsTHREEtsfamp}{\Large}
\newcommand{\lwTHREEtsfamp}{1.\linewidth}

\newcommand{\lwTWOLTHREEtsfamp}{1.\linewidth}

\newcommand{\cutA}{10}
\newcommand{\cutB}{30}
\newcommand{\cutC}{50}

\usepackage{siunitx} %
\newcommand{\numOf}[1]{\num[round-mode=figures,round-precision=1]{#1}}

\newcommand{\numEN}[1]{\num{#1}}

\newcommand{\jfmCBlabel}{$\times\mhelinj$}
\newcommand{\jfmCBEElabel}{$\times\EMinj$}

\begin{document}

\maketitle

\begin{abstract}

The role of the different helical components of the magnetic and velocity fields in the inverse spectral transfer of magnetic helicity is investigated through Fourier shell-to-shell transfer analysis. Both magnetic helicity and energetic transfer analysis are performed on chosen data from direct numerical simulations of homogeneous isothermal compressible magnetohydrodynamic turbulence, subject to both a large-scale mechanical forcing and a small-scale helical electromotive driving. The root mean square Mach number of the hydrodynamic turbulent steady-state taken as initial condition varies from 0.1 to about 11. Three physical phenomena can be distinguished in the general picture of the spectral transfer of magnetic helicity towards larger spatial scales: local inverse transfer, non-local inverse transfer and local direct transfer. A shell decomposition allows to associate these three phenomena with clearly distinct velocity scales while the helical decomposition allows to establish the role of the different helical components and the compressive part of the velocity field on the different transfer processes. The locality and relative strength of the different helical contributions are mainly determined by the triad helical geometric factor. 

\end{abstract}

\section{Introduction}

In three-dimensional turbulent hydrodynamic systems, the kinetic energy cascades to ever smaller scales through the self-similar break-up of eddies in smaller ones, which successively are subject to this self-similar process. As such, turbulence continuously transforms large-scale coherence into smaller scale structures which are eventually dissipated by viscous effects. Contrary to this intuition, turbulence in 3D magnetohydrodynamics (MHD) inherently generates large-scale magnetic structure, as well. One fundamental mechanism which is relevant in this perspective is the inverse spectral transfer (from large to small wavenumbers) of an ideal quadratic invariant, the magnetic helicity $\mhel=\volavg{\FmagA \cdot \Fmag}$, with $\FmagA$ the magnetic vector potential, $\Fmag = \nabla \times \FmagA$ the magnetic field and $\volavg{\cdot}$ denoting the volume average. In astrophysical systems, where the MHD single-fluid approximation is often satisfactory for the nonlinear dynamics of ionised gases on large scales, the resistivity is typically very low so that rotational motion leads naturally to helical magnetic fields \cite{ALF42}. Magnetic helicity dynamics are thought to play a crucial role in solar flares and coronal mass ejections \cite{LOW94,KUR96} while the solar wind is also associated with magnetic helicity transport \cite{BEM87,BER99}. \corr{Its dynamics are also relevant for plasma confinement in reversed-field-pinch fusion experiments \cite{EMO00} and its conservation plays e.g. a very important role in dynamo processes \cite{VIC01,BRA01,BRL13}.}

The inverse transfer of magnetic helicity has first been suggested by absolute-equilibrium statistical models \cite{FPL75} and successively has been verified by numerical experiments \cite{PFL76,POP78,MFP81,BAP99,CHB01,BRA01,AMP06,MAL09,MMB12,LBM16,LSM17}. In the present work, the terminology ``inverse transfer'' is preferred to ``inverse cascade'', since the word ``cascade'' is usually associated with a local transfer in Fourier space, involving fluctuations characterised by wavenumbers $\vk$, $\vp$, and $\vq$ with similar moduli. As shown here and in previous research by shell-to-shell transfer analysis \cite{AMP06}, both local and non-local aspects are present in this inverse spectral transport. Furthermore, using a decomposition of the fields along eigenvectors of the curl operator (``helical decomposition''), it has been shown that the inverse transfer of magnetic helicity in a wavevector triad, $\vk+\vp+\vq=\mathbf{0}$, is stronger and more non-local when the three interacting magnetic and velocity helical modes have the same helical sign \cite{LSM17}.

The above-mentioned studies \cite{AMP06,LSM17} have been performed in the incompressible case. This approximation, which is usually chosen for the sake of simplicity, is however of limited applicability for many natural and mainly astrophysical flows, which often are highly supersonic. For example, the root mean square (RMS) turbulent Mach number varies typically between 0.1 and 10 in the interstellar medium (\cite{ELS04}, section 4.2). In the present work, both the shell-to-shell transfer analysis and the helical decomposition are combined to study the inverse transfer of magnetic helicity in compressible isothermal ideal MHD turbulence. The aim is to disentagle the role of the different helical components of the magnetic and velocity fields by analysing data from direct numerical simulations of large-scale-mechanically-driven turbulence, with either a purely solenoidal or purely compressive forcing, and Mach numbers ranging from 0.1 to about 11. In these flows, small-scale helical magnetic fluctuations are injected.

The numerical experiments considered are described in section \ref{sec:numexp}. Sections \ref{sec:shelltoshell} and \ref{sec:helicaltransfers} describe the analytical tools used, namely the shell-to-shell transfer analysis and the helical decomposition, respectively. The results are reported in section \ref{sec:results}, while section \ref{sec:conclusion} summarises the findings and gives some concluding remarks.

\section{Numerical experiments}
\label{sec:numexp}

The numerical data is generated by solving the isothermal compressible ideal single-fluid MHD equations, in the presence of both a mechanical and an electromotive driving. \corrT{They read, in conservative form:}

\beqa
\pat \rho &=& - \nabla \cdot (\rho \Fvel),\\
\pat (\rho \Fvel) &=& - \nabla \cdot \left( \rho \Fvel \Fvel^T + (\rho \cs^2 + \frac{1}{2}|\Fmag|^2)\mathI - \Fmag \Fmag^T \right)+\rho\ForceV,\\
\label{eq:dtb} \pat \Fmag &=& \nabla \times (\Fvel \times \Fmag)+\ForceM,\\
\nabla \cdot \Fmag&=&0,
\eeqa

where $\rho$ is the mass density, $\Fvel$ the velocity and $\Fmag$ the magnetic field. In the isothermal case, the sound speed $\cs$ is constant, giving the (thermal) pressure $p=\rho \cs^2$. The $3\times 3$ identity matrix is denoted by $\mathI$. The driving terms $\ForceV$ and $\ForceM$ inject kinetic energy at large scales and small scale magnetic helical fluctuations respectively and are described below. The equations are solved using a fourth-order shock-capturing finite-volume solver, which makes use of the constrained-transport approach to ensure the solenoidality of the magnetic field. It is described in detail in \cite{TEI20,VTH19}. The main reconstruction method used is a fourth-order Central Weighted Essentially Non-Oscillatory (CWENO) method \cite{LPR99}, with a passage through point values in order to keep fourth-order accuracy \cite{COC11,BUH14}. The Riemann problems at the cell interfaces are solved using the Rusanov approximation \cite{RUS61} and a multidimensional version of the same is used for the line-integrated electric field in the constrained-transport framework \cite{BAL10}, which inherently maintains the solenoidality of the magnetic field up to machine precision \cite{EVH88}. The time-stepping is done through a Strong Stability Preserving Runge-Kutta (SSPRK) method, namely the one described by pseudocode 3 in \cite{KET08}, \corrT{with a timestep limited by the Courant-Friedrichs-Lewy criterion with a Courant number $C_{CFL}=1.5$}. In order to prevent the appearance of negative mass densities as a result of numerical inaccuracies at strong discontinuities and shocks prevalent in high Mach number flows, a local reduction of the scheme's order is used in the vicinity of discontinuities, a technique often referred to as ``flattening'' or ``fallback approach''. The increased accuracy of this higher-order numerical model permits to obtain results comparable to a standard second-order accurate numerical model at a significantly reduced resolution of the numerical grid \cite{VTH19}.

	The numerical experiments are performed as follows. Hydrodynamic turbulent steady-states are first generated from a fluid at rest ($\Fmag=\Fvel=0$, $\rho=\rho_0=1$) in $512^3$ cubic simulation boxes of size $\Lbox=1$ with triply periodic boundary conditions by injecting large-scale kinetic energy through an acceleration field $\ForceV$. The isothermal sound speed is $\cs=0.1$. The mechanical driving is carried out similarly to the experiments in \cite{FRD10,FED13}. It is governed by an Ornstein-Uhlenbeck process, which injects either purely solenoidal or purely compressive energy at the wavenumber shells $1 \leq \kspec \leq 2$ (with the shell $\kspec \in \mathbb{N}$ defined by the wavevectors $\vk$ such that $\kspec \leq |\vk|/\kone < \kspec+1$, with $\kone=2\upi/\Lbox$ the smallest wavenumber in the system). The energy injection rate $\Einj$ governs the turbulent RMS Mach number $\rmsM$ of the statistical steady-state obtained when numerical dissipation balances the injected energy. The forcing auto-correlation time is roughly the turbulent turnover time $\ttt=\Lbox/(2\cs\rmsM)$. The weak mean velocity field which appears as a result of the forcing is removed at each iteration. When the steady-state is reached, a delta-correlated electromotive driving $\ForceM$ is switched on at a particular instant, which injects fully helical magnetic fluctuations at small scales $\kspec \in [48,52]$ with a defined magnetic energy injection rate $\EMinj$. This means that only one sign of magnetic helicity is injected, which dominates the system. This simplification allows to limit the complexity of the present study to a practical level.

	In the present paper, shell-to-shell helical analysis of the least and most compressible runs presented in \cite{TEI20} are considered, which are labelled by ``M01s4'', ``M11s'', ``M1c'' and ``M8c''. The number in the label stands for the approximate RMS Mach number during the hydrodynamic turbulent steady-state, which is $\rmsM \approx 0.116, 11.1, 0.797$ and $7.87$ respectively, and the letter for the forcing type, either purely solenoidal or compressive. The magnetic-to-energy injection rate $\EMinj/\Einj$ is taken as unity for all the runs but the M01s4 one for which $\EMinj/\Einj=4$, a value that has been chosen in order to obtain faster convergence of spectral scaling laws, an aspect which is not studied here.

\section{Shell-to-shell transfers}
\label{sec:shelltoshell}
	The formalism used in the incompressible case in \cite{AMP05,AMP06} for magnetic helicity and energetic shell-to-shell transfers is reviewed and extended here for compressible MHD. For a field $\neutrfield$, the field obtained by keeping only the wavenumbers in a certain shell $\kspec \in \mathbb{N}$, defined by the wavevectors $\vk$ such that $\kspec \leq |\vk|/\kone < \kspec+1$ (with $\kone=2\upi/\Lbox$ the smallest wavenumber in the system) is labelled by $\neutrfield_K$.

\subsection{Magnetic helicity transfers}

	In the absence of forcing and of dissipative effects, \corrT{$\mhelFK=\volavg{\FmagAK \cdot \FmagK}$, the magnetic helicity present in shell $K$}, is governed by \cite{AMP06}:

\beq
	\label{eq:tsfmhlqk}
	\pat \mhelK = \sum_Q \TsfMhlQK= \sum_Q 2\volavg{\FmagK \cdot (\Fvel \times \FmagQ)},
\eeq

	where $\TsfMhlQK$ corresponds to the transfer rate of magnetic helicity from shell $Q$ to shell $K$. This transfer function, derived in the incompressible case, remains valid without modification in compressible MHD. The interpretation as a transfer rate of magnetic helicity between shells is justified through the antisymmetric property $\TsfMhlQK=-\TsfMhl(K,Q)$. Please note that in \cite{AMP06}, the inverse convention is used, with $T_h(Q,K)$ the transfer of magnetic helicity received by shell $Q$ from shell $K$. The magnetic helicity is also defined there as $\mhel=\frac{1}{2} \int \FmagA \cdot \Fmag \dvol$ so that a factor 2 is not present in the transfer rates. The velocity field can also be decomposed in shells, yielding the transfer function:

\beq
	\label{eq:tsfmhl}
	\TsfMhlQPK= 2\volavg{\FmagK \cdot (\FvelP \times \FmagQ)},
\eeq

which represents the transfer of magnetic helicity from shell $Q$ to shell $K$ mediated by the velocity field at shell $P$. The structure of relation \eqref{eq:tsfmhl} reflects that magnetic helicity transfers occur indeed through triad interactions, even in compressible MHD (see section \ref{sec:triaddef}). Since magnetic helicity is a purely magnetic ideal invariant, the velocity field cannot exchange helicity with the magnetic field and hence plays only a mediating role for transfers of magnetic helicity between two magnetic modes. \corr{Summing the function of three variables $\TsfMhlQPK$ along $Q$ allows to quantify the importance of the mediating velocity field with respect to magnetic helicity transfers to shell $K$ as a function of the scale $P$:}

\beq
	\label{eq:MhlPK}
	\corr{\TsfMhlPK=\sum_Q \TsfMhlQPK}
\eeq

	In this work, the indices $K$ and $Q$ always correspond to magnetic field shells, whereas the index $P$ always corresponds to the velocity field. 

\subsection{Energetic transfers}
\label{sec:enetsf}

For shell-to-shell energetic transfers, the approach in \cite{AMP05} needs to be extended by considering the influence of the magnetic pressure term. In compressible isothermal MHD, energy can be stored in three reservoirs: as kinetic energy $\Ekin=\frac{1}{2}\rho\Fvel^2$, as magnetic energy $\Emag=\frac{1}{2}\Fmag^2$ and as potential energy, in the form of density fluctuations $\epot=\rho \cs^2 \ln(\rho/\rhoz)$, with $\rhoz$ the mean density in the system \cite{KWN13}. These energy forms are governed by:

	\beqa
                \pat \ekin &=& \underbrace{- \nabla \cdot (\frac{1}{2} \rho |\Fvel|^2 \Fvel)}_{Ia} + \underbrace{\Fvel \cdot ((\Fmag \cdot \nabla) \Fmag) - \Fvel \cdot \nabla (\frac{1}{2}|\Fmag|^2) }_{Ib} \underbrace{- \Fvel \cdot \nabla (p)}_{Ic},\\
                \pat \emag &=& \underbrace{- \Fmag \cdot ((\Fvel \cdot \nabla) \Fmag)- \frac{1}{2} |\Fmag|^2 \nabla \cdot \Fvel}_{IIa} + \underbrace{\Fmag \cdot ((\Fmag \cdot \nabla) \Fvel)- \frac{1}{2} |\Fmag|^2 \nabla \cdot \Fvel}_{IIb},\\
                \pat \epot&=&- \nabla \cdot (p \Fvel + p \Fvel \ln(\rho)) + \underbrace{\Fvel \cdot \nabla (p)}_{IIIc}.
        \eeqa

	The term $IIa$ can be rewritten in divergence form $\nabla \cdot (\ldots)$, so that the $Ia$ and $IIa$ terms are conservative and correspond to energy exchanges between scales within one field (kinetic or magnetic). The other terms with underbraces correspond to energy exchanges between reservoirs: the $Ib$ and $IIb$ terms to energy exchanges between the kinetic energy and the magnetic energy reservoirs (since $\Fvel \cdot ((\Fmag \cdot \nabla) \Fmag) - \Fvel \cdot \nabla (\frac{1}{2}|\Fmag|^2)=\nabla \cdot ((\Fvel \cdot \Fmag) \Fmag - \frac{1}{2} |\Fmag|^2 \Fvel) - \Fmag \cdot ((\Fmag \cdot \nabla) \Fvel) + \frac{1}{2} |\Fmag|^2 \nabla \cdot \Fvel$)
 whereas the $Ic$ and $IIIc$ terms correspond to energy exchanges between the potential energy and the kinetic energy reservoirs. This leads to the following transfer functions:

\beqa
	\label{eq:tsfekbv}
	\TsfEkBVKPcomp&=& \volavg{\FvelP \cdot (\Fmag \cdot \nabla) \FmagK - \frac{1}{2} (\FvelP \cdot \nabla) (\Fmag \cdot \FmagK)},\\
	\label{eq:tsfembb}
	\TsfEmBBQKcomp&=& \volavg{-\FmagK \cdot (\Fvel \cdot \nabla) \FmagQ - \frac{1}{2} (\FmagK \cdot \FmagQ) \nabla \cdot \Fvel},\\
	\label{eq:tsfemvb}
	\TsfEmVBPKcomp&=& \volavg{\FmagK \cdot (\Fmag \cdot \nabla) \FvelP - \frac{1}{2} (\FmagK \cdot \Fmag) \nabla \cdot \FvelP},
\eeqa

	with $\Tsf^{\neutrenergysymb^j}_{xy}(S,T)$ the energy transfer from shell $S$ of field $x$ to shell $T$ of field $y$ (with $x,y \in \{v,b\}$). The superscript $\neutrenergysymb^j$, corresponding to field $y$, is thus redundant but kept for clarity reasons. The antisymmetric properties $\TsfEmBBQKcomp=-\TsfEmBBKQcomp$ and $\TsfEmVBPKcomp=-\TsfEkBVKPcomp$ justify the interpretation of relations \eqref{eq:tsfekbv}-\eqref{eq:tsfemvb} as energetic transfers between reservoirs and scales. The $\TsfEmVBcomp$ term can be further decomposed in contributions coming from the magnetic stretching $\TsfEmKBT$ and magnetic pressure $\TsfEmKBP$ terms, defined as:

\beq
	\label{eq:magpresstretch}
	\TsfEmKBTPK = \volavg{\FmagK \cdot (\Fmag \cdot \nabla) \FvelP}, \;\;\; \TsfEmKBPPK= \volavg{-\frac{1}{2} (\FmagK \cdot \Fmag) \nabla \cdot \FvelP}.
\eeq

The energetic shell-to-shell transfers among the velocity fluctuations as well as those between the kinetic and the potential energy reservoirs are not considered in this work. Care has to be taken when interpreting the exchanges between the magnetic and kinetic reservoirs: since the mass density is not constant in compressible MHD, $\TsfEmVBPKcomp$ does \textit{not} represent the transfer of kinetic energy at shell $P$ to magnetic energy at shell $K$, but only kinetic energy \textit{associated} with $\FvelP$ to magnetic energy at shell $K$. A possibility to express the kinetic energy in shell $P$ is $\volavg{|\rhowIIv_P|^2/2}$ with $\rhowIIv=\sqrt{\rho}\Fvel$. Transfer functions using this approach in compressible MHD have been derived in \cite{GOS17}. \corr{They are not used here for reasons of conceptual simplicity: the helical decompositions of $\Fvel$ and $\Fmag$ are indeed easier to interpret than those of $\rhowIIv$ and $\FvelA=\Fmag/\sqrt{\rho}$.}

	\corr{Furthermore, the impact of energetic exchanges on the magnetic helicity shell-to-shell transport can be determined by decomposing relation \eqref{eq:tsfmhlqk} in $\TsfMhlQK=\TsfMhlBBQK+\TsfMhlKBTQK+\TsfMhlKBPQK$ with:}
{
\corr{
\beqa
	\label{eq:tsfmhlEneA}
	\TsfMhlBBQK&=& 2\volavg{\FmagK \cdot \uncurl( (-\Fvel \cdot \nabla) \FmagQ - \frac{1}{2}  \FmagQ \nabla \cdot \Fvel)},\\
	\label{eq:tsfmhlvbqk}
	\TsfMhlKBTQK&=&2\volavg{\FmagK \cdot \uncurl( (\FmagQ \cdot \nabla) \Fvel)},\\
	\label{eq:tsfmhlEneB}
	\TsfMhlKBPQK&=& 2\volavg{\FmagK \cdot \uncurl( - \frac{1}{2} \FmagQ \nabla \cdot \Fvel)},
\eeqa
}
}

\corr{where $\uncurl(\neutrfield)$ returns the solenoidal field $\neutrfieldB$ whose curl is the solenoidal part of field $\neutrfield$ (defined as $\neutrfieldBF_{\vk}=\frac{i \vk \times \neutrfieldF_{\vk}}{k^2}$ in Fourier space). These functions have been obtained by appropriate ``transcription'' of relations \eqref{eq:tsfembb} and \eqref{eq:magpresstretch} in terms of magnetic helicity transfers. Contrary to their sum, the $\TsfMhlBBQK$, $\TsfMhlKBTQK$ and $\TsfMhlKBPQK$ functions cannot be on their own interpreted as magnetic helicity shell-to-shell transfers since they are not antisymmetric.}

\corrT{In order to relate more easily to physical quantities, the Fourier spectra shown in the present work are normalised by the isothermal sound speed squared, i.e., the specific kinetic energy spectrum is $\sekinFK=\frac{1}{2\cs^2}\volavg{\Fvel^2_K}$, the magnetic energy $\emagFK=\frac{1}{2\rho_0\cs^2}\volavg{\Fmag^2_K}$ with $\rho_0=1$ the mean density in the system and similarly $\mhelFK=\frac{1}{\rho_0\cs^2}\volavg{\FmagA_K \cdot \Fmag_K}$.}

\section{Helical transfers}
\label{sec:helicaltransfers}

	The helical decomposition is based on the diagonalisation of the curl operator in Fourier space $\neutrfieldF \to i \vk \times \neutrfieldF$, which possesses the eigenvalues $(0,+k,-k)$ (with $k=|\vk|$) associated respectively with the unitary eigenvectors $\vhhk = \vk/k$ and \cite{WAL92,BRS05}:

\beq
		\vhhpm = \frac{1}{\sqrt{2}}\frac{\vk \times (\vk \times \vhe) \mp i k (\vk \times \vhe)}{k^2\sqrt{1-(\vk \cdot \vhe/k)^2}},
\eeq

with $\vhe$ an arbitrary vector of unitary length non-parallel to $\vk$. This decomposition can be viewed as an extension of the Helmholtz decomposition in a solenoidal (orthogonal to $\vk$) and a compressive (parallel to $\vk$) part. The plane orthogonal to $\vk$ is spanned by $(\vhhp,\vhhm)$ so that the solenoidal modes are further decomposed in circularly polarised waves with opposite polarity, corresponding in configuration space to flow lines forming either a right-handed or a left-handed helix. The study of helical mode interactions allows to find interactions responsible for a sub-dominant inverse transfer of kinetic energy in 3D hydrodynamic turbulence \cite{WAL92,ALE17} and extensions to the MHD case seem promising to clarify the intertwined dynamics of magnetic and kinetic helicities \cite{LPC09,LSM17,ALB18}.
	Using this decomposition, the velocity and magnetic fields read \cite{LPC09}:

\beqa
	\label{eq:heldecompB}
	\Fmag&=&\sum_{\vk} \sum_{\hsk} \FmagFL^{\hsk}_{\vk} \vhhs{\hsk}{\vk} e^{i\vk \cdot \vx},\\
	\label{eq:heldecompV}
	\Fvel&=&\sum_{\vp} \sum_{\hsp} \FvelFL^{\hsp}_{\vp} \vhhs{\hsp}{\vp} e^{i\vp \cdot \vx},
\eeqa

	with $\hsk \in \{+,-\}$ corresponding to the positively and negatively helical modes and $\hsp \in \{+,-,0\}$, corresponding to both helical modes and the compressive mode along $\vk$. Since the magnetic field is solenoidal, it cannot have components along $\vhhk$.

Plugging the helical decomposition \eqref{eq:heldecompB}-\eqref{eq:heldecompV} in \eqref{eq:dtb} in the absence of forcing leads to \cite{LPC09,LSM17}:

\beq
	\label{eq:pathelb}
	\pat \FmagFL^{\hsk}_{\vk}=\hsk k \sum_{\vk+\vp+\vq=0} \sum_{\hsp,\hsq} \FvelFL^{\hsp*}_{\vp} \FmagFL^{\hsq*}_{\vq} \triadgeomkpq,
\eeq

\label{sec:triaddef}
with $\triadgeomkpq$ a geometric factor depending on the triad shape and the helical components considered. Its modulus $G^{\hsk,\hsp,\hsq}_{k,p,q}$ can be viewed as a weight of the respective helical triad interaction. The derivations performed in incompressible MHD in \cite{LPC09} remain valid in the compressible case, with the only difference that the index for the velocity helical component can also correspond to the compressive mode ($\hsp=0$).
	The geometric factor has been derived in \cite{WAL92} for triads in the incompressible case. It is also valid in the compressible case when $\hsp \in \{+,-\}$, with its modulus given by:

\beq
	\label{eq:Gincomp}
	G^{\hsk,\hsp \in \{+,-\},\hsq}_{k,p,q}=\frac{|\hsk k+\hsp p+\hsq q|\sqrt{2k^2p^2+2p^2q^2+2q^2k^2-k^4-p^4-q^4}}{2kpq}.
\eeq

	\corrT{These derivations are repeated in appendix \ref{app:Qgeom}. An extension to compressible flows is also presented there, leading for $\hsp=0$ and $\hsk=S$, $\hsq=\pm S$ to:}

\beq
	\label{eq:Gcomp}
	G^{\hsk=S,\hsp=0,\hsq=\pm S}_{k,p,q}=\frac{|(q\mp k)(p^2-k^2-q^2 \mp 2qk)|}{2kpq}.
\eeq

\label{sec:helstos}
	The shell-to-shell transfer approach described in section \ref{sec:shelltoshell} can be extended to helically-decomposed shell-to-shell transfers. The magnetic field is projected in Fourier space on the helical eigenvectors $\vhhp$ and $\vhhm$ and then transformed into configuration space, giving $\FmaghP$ and $\FmaghN$ respectively (where ``$P$'' and ``$N$'' stand for ``positively'' and ``negatively'' helical). In the same way, the velocity field is decomposed in $\FvelhP$, $\FvelhN$ and $\FvelhC$, with $\FvelhC$ its compressive part (obtained through a projection along $\vhhk$ in Fourier space). The magnetic helicity transfer function (relation \eqref{eq:tsfmhl}) is then extended to:

\beq
	\TsfMhlQPK=\sum_{\hsK \in \{\helP,\helN\}}\sum_{\hsP \in \{\helP,\helC,\helN\}}\sum_{\hsQ \in \{\helP,\helN\}} \TsfMhlXXX{\hsK\hsP\hsQ},
\eeq

	with twelve different helical contributions:

\beq
	\TsfMhlXXX{\hsK\hsP\hsQ}(Q,P,K)=2\volavg{\FmagXX{\hsK}{K} \cdot (\FvelXX{\hsP}{P} \times \FmagXX{\hsQ}{Q})}.
\eeq

	The shell-to-shell energetic transfer functions (section \ref{sec:enetsf}) are extended analogously. The six helical transfer functions for which $\hsK=\hsQ$ can be interpreted as antisymmetric magnetic helicity transfers between shells $\TsfMhlXXX{\hsK\hsP\hsK}(Q,P,K)=-\TsfMhlXXX{\hsK\hsP\hsK}(K,P,Q)$. These terms are labelled ``$\PPP, \PNP, \PCP, \NPN, \NNN$'' and ``$\NCN$'' in the following, where the first letter corresponds to $\hsK$, the second to $\hsP$ and the third to $\hsQ$. However, the terms where $\hsK \neq \hsQ$, labelled henceforth as ``heterochiral'' are not antisymmetric on their own but need to be considered in pairs, as in \cite{ALE17}: the three ``(s)ymmetrised'' terms $\TsfMhlXXX{\helS\hsP\helS}=\TsfMhlXXX{\helP\hsP\helN}+\TsfMhlXXX{\helN\hsP\helP}$ verify the antisymmetric property and can thus be interpreted as magnetic helicity transfers between shells. They are noted similarly ``$\SPS, \SNS$'' and ``$\SCS$''. 

	Even though the triad helical geometric factor $G^{\hsk,\hsp,\hsq}_{k,p,q}$ is strictly speaking only valid for single triad interactions, one can expect the helically-decomposed shell-to-shell transfers to be weighted essentially by the same geometric factor, especially for $K, P, Q$ large enough.

\section{Results}
\label{sec:results}

 \begin{figure}%
     \centering
	\vspace{\plotvspaceval em}
     \huge
     \resizebox{0.6\linewidth}{!}{\input{\grpathstos_spec_jfm.tex}}

	\vspace{\plotvspacevalend em}

     \caption[Small caption]{\footnotesize{Magnetic helicity $\mhelF$, magnetic energy $\emagF$ and specific kinetic energy $\sekinF$ spectra for the M8c run at an instant when $\IscHm \approx \Lbox/10$.}}
     \label{fig:scalspec}
 \end{figure}

	The helically-decomposed shell-to-shell transfer functions are considered for the least and most compressible runs M01s4, M11s, M1c and M8c at the respective instant in time where the magnetic helicity integral scale $\IscHm=\Lbox(\int_{\kspec} \kspec^{-1} \mhelF d\kspec)/(\int_{\kspec} \mhelF d\kspec) \approx \Lbox/10$. Even though magnetic helicity is not sign-definite, since only one helical sign is injected, positive magnetic helicity dominates the system at all scales so that considering its integral scale is meaningful. \corrT{This instant is chosen so that self-similar dynamics are observed while the spectral pollution by large-scale condensation remains negligible, as shown through the spectra displayed in figure \ref{fig:scalspec}. Spectral scaling laws are explored in a forthcoming paper.}

	In the following, only the most compressible M8c run is analysed in greater detail, because it is an extreme and clear realisation of similar inverse transfer dynamics, which have been observed in previous research only at lower Mach numbers or in the incompressible case \cite{BAP99,CHB01,BRA01,MMB12}. A short comparison with the other runs is made in section \ref{sec:results_otherruns}. \corr{The plots are in units of the (estimated) magnetic helicity injection rate, $\mhelinj=(2\EMinj\mhelfracFor)/(2\upi\mhlkinj/\Lbox)$, with $\mhelfracFor=1$ the helical fraction of the injected fluctuations and $\mhlkinj=50$ the shell around which magnetic helicity is injected.}

	Figure \ref{fig:tsfamp_hm_M7c} shows the general aspect of the shell-to-shell magnetic helicity transfers $\TsfMhlQK$ for the M8c run. As expected, this figure is antisymmetric about the main diagonal. A positive value (yellow/bright colour) above the diagonal means a positive transfer from shell $Q<K$ to shell $K$, that is a direct transfer, whereas a negative value (blue/dark colour) above the diagonal means an inverse transfer, and vice-versa for the values below the diagonal. The farther from the diagonal, the more distant the shells $K$ and $Q$ are, which indicates the transfers' level of locality. Since a linear binning of the shells is chosen here, care has to be taken when interpreting the figures, since such a binning may lead to an overestimation of the transfers' non-locality \cite{AEY10,GOS17}. A linear binning is chosen here in order to see better the different features, since a logarithmic binning would leave only few shells. For visualisation purposes also, the colour bar extremes are smaller than the extremes of the transfer functions. The aim of the 2D plots is indeed to present a qualitative behaviour regarding the strength, direction and locality of the transfers. Cuts along one direction are shown when a more quantitative aspect needs to be underlined.

\def\plotvspaceval{-2}
\def\plotvspacevalend{-2}
\def\tsPlotOne{\small}
\def\tsPlotTwo{\Large}
	 
 \begin{figure}
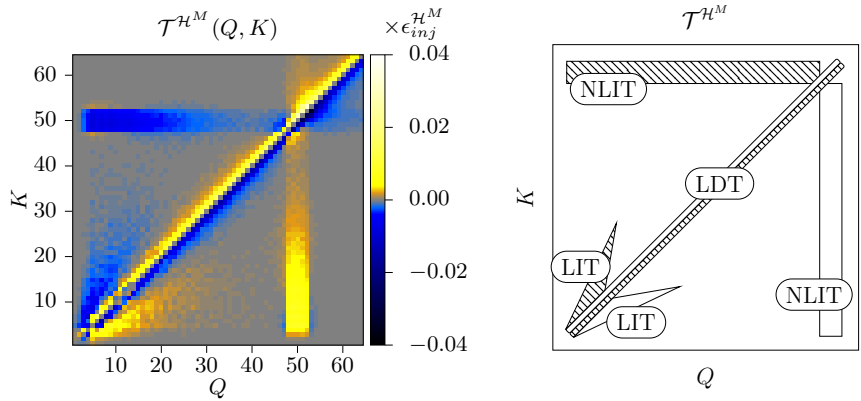
%
     \centering
	\vspace{\plotvspaceval em}
\begin{minipage}{0.59\linewidth}
	\tsPlotOne
 	\resizebox{\textwidth}{!}{\input{\grpathtsfamp_hm_M7c10.tex}}
\end{minipage}%
\begin{minipage}{0.4\linewidth}
	\tsPlotTwo
	\resizebox{\textwidth}{!}{\input{\grpathsketch_tsfamp.tex}}
\end{minipage}

	\vspace{\plotvspacevalend em}

     \caption[Small caption]{\footnotesize{Left: magnetic helicity transfer rates from shell $Q$ to shell $K$ for the \MVIIc\ run, at an instant when $\spectransMainInstant$. Right: a sketch of these transfer rates expliciting the non-local inverse transfer (NLIT), the local inverse transfer (LIT) and the local direct transfer (LDT).}}

     \label{fig:tsfamp_hm_M7c}
 \end{figure}

\def\plotvspaceval{0}
\def\plotvspacevalend{0}

	Three phenomena can thus be distinguished in the general picture of a magnetic helicity inverse transfer as sketched in figure \ref{fig:tsfamp_hm_M7c}, right: a local inverse transfer (corresponding to the ``wings'' in the lower left corner close to the diagonal, henceforth named ``LIT''), a non-local inverse transfer (corresponding to vertical and horizontal stripes, hereafter ``NLIT'') and a local direct transfer (along the diagonal, ``LDT'' in the following). The main aim of the next subsections is to shed some light on their respective origin.

\subsection{Role of the mediating velocity field at different scales}
\label{sec:results_mediator}

\begin{figure}%
     \centering
	\vspace{\plotvspaceval em}
     \tsdefault
     \resizebox{\lwTWOLTHREEtsfamp}{!}{\input{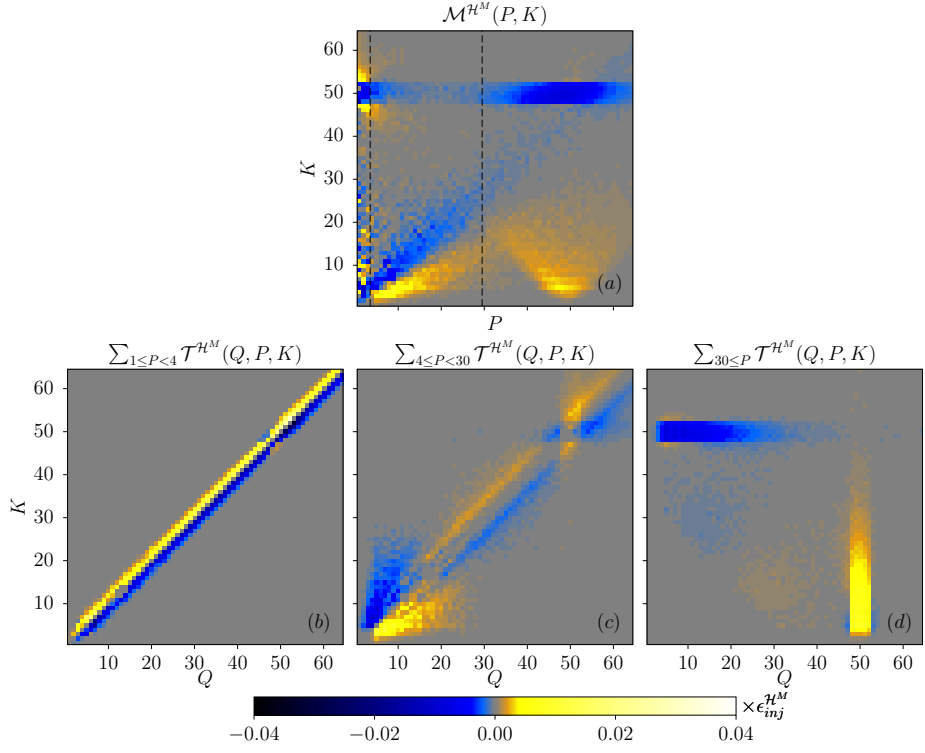}}

	\vspace{\plotvspacevalend em}

     \caption[Small caption]{\footnotesize{Role of the mediating velocity field, M8c run at an instant when $\IscHm \approx \Lbox/10$. $(a)$ Sum of $\TsfMhlQPK$ along $Q$. $(b\sm d)$ Sum of $\TsfMhlQPK$ along $P$ for different velocity field shells, showing that the three features are associated with mediation at different scales.}}
     \label{fig:tsfamp_hm_M7c_Vfil}
 \end{figure}

	Figure \ref{fig:tsfamp_hm_M7c_Vfil} shows that each of the LDT, LIT and NLIT feature can be associated with different velocity scales. The function $\TsfMhlPK$ (relation \ref{eq:MhlPK}), which quantifies the importance of the mediating velocity field at shell $P$ with respect to transfers of magnetic helicity to shell $K$ allows to delimitate three different regions along the $P$-axis (figure \ref{fig:tsfamp_hm_M7c_Vfil}.$(a)$). The small-scale velocity field (defined as $30 \leq P$) transfers magnetic helicity from the electromotive driving scale to significantly larger scales, and corresponds hence to the NLIT. This process can be interpreted as the spectrally nonlocal merging of small-scale magnetic fluctuations with a magnetic structure of much larger size. The intermediate-scale velocity field (defined as $4 \leq P < 30$) transports magnetic helicity from shells \corrT{$5 \lessapprox K \lessapprox 25$ to larger scales which are quite close spectrally (within a factor two)}. This allows the interpretation of merging like-sized magnetic fluctuations and corresponds hence to the LIT. Finally, the large-scale velocity field ($1\leq P<4$), with alternating signs along the $K$-axis is associated with the LDT. Along the same line of interpretation, the LDT would correspond to the effect of the direct cascade of kinetic energy, i.e. the destruction of magnetic structures by advective shear exerted by the velocity field. The separation of transfer dynamics in Fourier space is confirmed on figures \ref{fig:tsfamp_hm_M7c_Vfil}.$(b\sm d)$, where the sum of $\TsfMhlQPK$ for $P$ corresponding to the three above-mentioned velocity scales is plotted.

\subsection{Helical components}
\label{sec:results_helical}
	Since only magnetic helicity of one sign is injected by the electromotive forcing, the dominant transfer terms are expected to be the ones involving the positively helical magnetic field, labelled $\PXP$ in section \ref{sec:helstos}, with $X \in \{P,N,C\}$ corresponding respectively to the positively (``like-signed'' in the context of dominant positive magnetic helicity) and negatively (``opposite-signed'') helical parts of the velocity field and its compressive part. These terms hence allow to assess the role of the velocity field's helical components, as well as the role of its compressive part.

	For the M8c run, both dominant $\PPP$ and $\PCP$ terms, mediated by the like-signed helical velocity field and its compressive part respectively, contribute to the NLIT in a similar way, but their contributions to the LDT and the LIT differ significantly (figure \ref{fig:tsfamp_hm_M7c_PPP_PCP}). The $\PCP$ term takes the leading role in the LDT, but does essentially not contribute to the LIT. A more quantitative view of these aspects is available through figure \ref{fig:tsfamp_hm_M7c_cuts_PPP_PCP}, where cuts of figure \ref{fig:tsfamp_hm_M7c_PPP_PCP} along different $K=\KZ$ are shown.

\def\plotvspaceval{0.5}
	 
 \begin{figure}%
     \centering
	\vspace{\plotvspaceval em}
     \tsTHREEtsfamp
     \resizebox{\lwTHREEtsfamp}{!}{\input{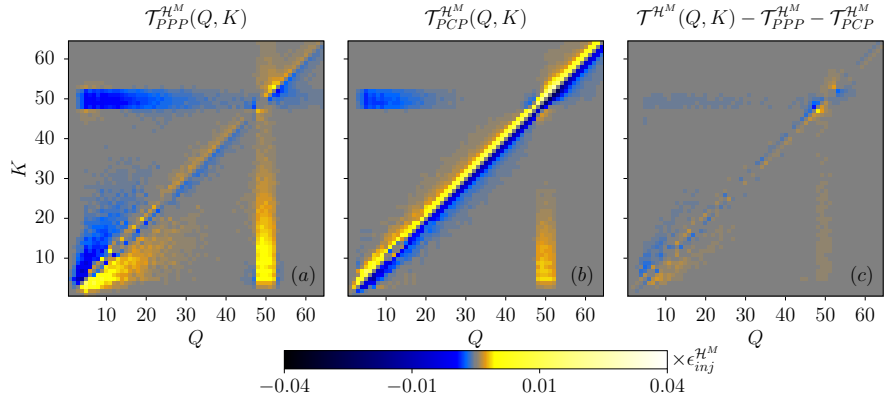}}

	\vspace{\plotvspacevalend em}

     \caption[Small caption]{\footnotesize{Contributions from the $\PPP$ and $\PCP$ terms to the magnetic helicity transfer rates for the \MVIIc\ run, at an instant when $\spectransMainInstant$. Their sum is close to the total transfer function from figure \ref{fig:tsfamp_hm_M7c} and the remaining contributions come mostly from the terms shown in figure \ref{fig:tsfamp_hm_M7c_PNP_NCN_SCS}.}}
     \label{fig:tsfamp_hm_M7c_PPP_PCP}
 \end{figure}

\def\plotvspaceval{0}

 \begin{figure}%
     \centering
	\vspace{\plotvspaceval em}
     \Huge
     \resizebox{1.\linewidth}{!}{\input{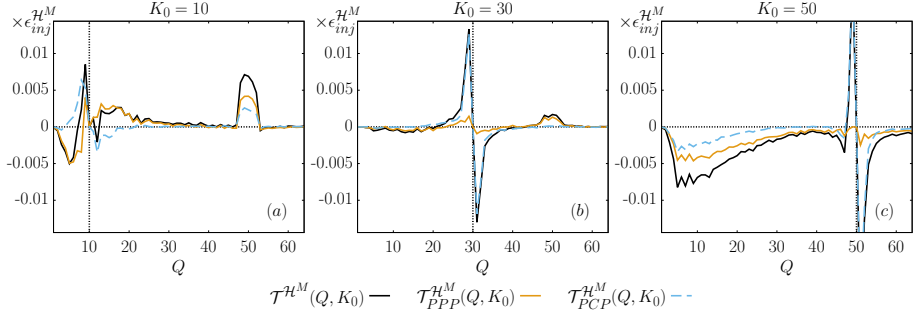}}

	\vspace{\plotvspacevalend em}

     \caption[Small caption]{\footnotesize{Cuts from the 2D plots of figure \ref{fig:tsfamp_hm_M7c_PPP_PCP} for $\KZ \in \{\cutA, \cutB, \cutC\}$. The horizontal dotted line corresponds to $y=0$ and the vertical one to $Q=\KZ$.}}
     \label{fig:tsfamp_hm_M7c_cuts_PPP_PCP}
 \end{figure}

 The other helical contributions are relatively small (figure \ref{fig:tsfamp_hm_M7c_PPP_PCP}.$(c)$). The three biggest remaining contributions, corresponding to the $\PNP$, $\NCN$ and $\SCS$ terms, are displayed in figure \ref{fig:tsfamp_hm_M7c_PNP_NCN_SCS}. The $\PNP$ term, mediated by the opposite-signed helical velocity field, presents a small contribution to the LIT and the LDT (please note that the colour bar extremes are an order of magnitude smaller than those of figure \ref{fig:tsfamp_hm_M7c_PPP_PCP}), and a minor contribution to the NLIT. The $\NCN$ term corresponds to a local direct transfer of \textit{negative} magnetic helicity, mediated by the compressive velocity field, which results in a local \textit{inverse} transfer of magnetic helicity. The heterochiral $\SCS$ term exhibits a more exotic shape. Its analysis is more complex since it is the sum of two terms corresponding to different geometric helical triad factors. It is not analysed in more details here since it corresponds to a relatively small contribution. The remaining $\NNN, \NPN, \SPS$ and $\SNS$ helical contributions are an order of magnitude smaller as compared to the $\PNP, \NCN$ and $\SCS$ terms and are not shown here.

\def\plotvspaceval{0.5}
	 
 \begin{figure}%
     \centering
	\vspace{\plotvspaceval em}
     \tsTHREEtsfamp
     \resizebox{\lwTHREEtsfamp}{!}{\input{\grpathtsfamp_hm_M7c10_PNP_NCN_SCS.tex}}

	\vspace{\plotvspacevalend em}

     \caption[Small caption]{\footnotesize{Contributions from the $\PNP$, $\NCN$ and $\SCS$ terms to the magnetic helicity transfer rates for the \MVIIc\ run, at an instant when $\spectransMainInstant$.}}
     \label{fig:tsfamp_hm_M7c_PNP_NCN_SCS}
 \end{figure}

\def\plotvspaceval{0}

	To summarise, for the M8c run, the LIT is exclusively mediated by the solenoidal part of the velocity field ($\PPP$ and $\PNP$ terms), with a significantly greater importance of the like-signed ($\PPP$) helical part. The velocity field's compressive part ($\PCP$ term) takes the leading role in the LDT, even though its solenoidal part contributes to it to a smaller extent. As for the NLIT, it is essentially mediated by both the like-signed velocity helical field and its compressive part, while the role of the opposite-signed helical velocity field is comparatively very small. The compressive part of the velocity field, which is not present in incompressible turbulence, enhances hence the direct magnetic energy cascade through shearing effects and the nonlocal merging of small-scale magnetic structures to structures of much larger size.

\subsection{Interpretation through the geometric factor}
\def\plotvspaceval{0.5}
\def\tsPlotTwo{\small}
\def\tsPlotOne{\Large}
	 
 \begin{figure}
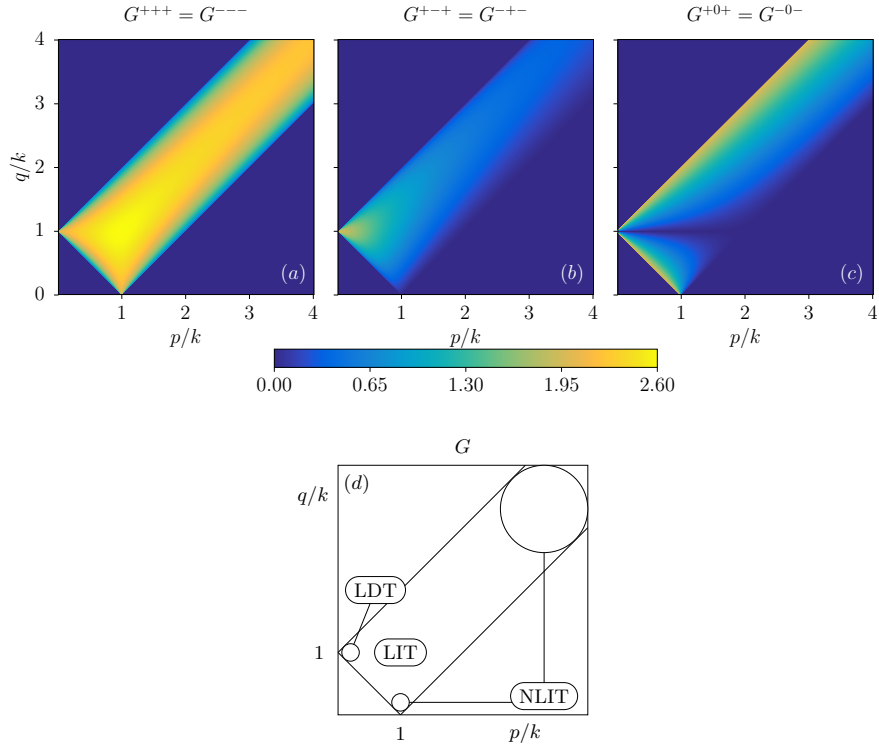
%
     \centering
	\vspace{\plotvspaceval em}
{
	\tsPlotOne
 	\resizebox{\lwTHREEtsfamp}{!}{\input{\grpathQgeom_homochiral.tex}}
}
	\vspace{0.5em}

{	\tsPlotTwo
	\resizebox{\lwTHREEtsfamp}{!}{\input{\grpathsketch_Qgeom_G.tex}}
}

	\vspace{\plotvspacevalend em}

     \caption[Small caption]{\footnotesize{$(a\sm c)$Triad helical geometric factor's modulus for the $\PPP$, $\PCP$ and $\PNP$ terms respectively (see relations \eqref{eq:Gincomp} and \eqref{eq:Gcomp}), $(d)$ Sketch showing to which regions the NLIT, the LIT and the LDT correspond.}}

     \label{fig:QgeomPKM}
 \end{figure}

\def\plotvspaceval{0.}

 \begin{figure}%
     \centering
	\vspace{\plotvspaceval em}
     \huge
     \resizebox{\lwTHREEtsfamp}{!}{\input{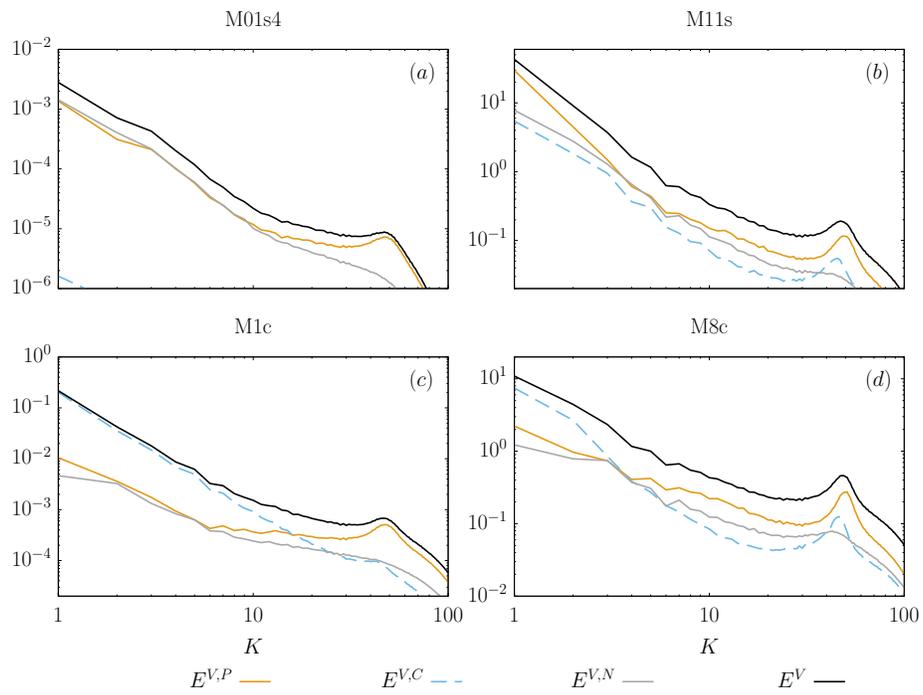}}

	\vspace{\plotvspacevalend em}

     \caption[Small caption]{\footnotesize{Repartition of the specific kinetic energy in the helical components for all the extreme runs, at an instant when $\IscHm \approx \Lbox/10$.}}
     \label{fig:EkVPKMcomp}
 \end{figure}

	\corr{The results from section \ref{sec:results_helical} regarding the role of the $\PPP$, $\PNP$ and $\PCP$ terms can be explained through the moduli $G^{+++}$, $G^{+-+}$ and $G^{+0+}$ of the respective helical geometric triad factors (relations \eqref{eq:Gincomp}-\eqref{eq:Gcomp}), displayed in figure \ref{fig:QgeomPKM}.} The sketch in that figure summarises the results from section \ref{sec:results_mediator}. The LDT, which is associated with the large-scale velocity field, corresponds to $p \ll k \approx q$, whereas the LIT, associated with the intermediate-scale velocity field, corresponds to $p \approx k \approx q$. As for the NLIT, it is associated with the small-scale velocity field, that is $k \ll p \approx q$ and $q \ll p \approx k$.

	The relative importance of the $\PPP$, $\PNP$ and $\PCP$ terms for the three NLIT, LIT and LDT features can be related to the geometric factor as follows:
	\begin{itemize}
		\item In the region $k \approx p \approx q$ corresponding to the LIT, the $G^{+++}$ term is biggest with a maximum at $3\sqrt{3}/2 \approx 2.6$ for $k=p=q$, whereas $G^{+-+}<G^{+++}$ by a factor of about 3 and $G^{+0+}$ vanishes in this region. This explains why the LIT is mediated by the like-signed helical velocity field ($\PPP$ term) and to a lesser extent by the opposite-signed helical velocity field ($\PNP$), but is essentially absent from the $\PCP$ term, corresponding to a mediation by the compressive velocity field.
		\item In the region corresponding to the NLIT, both $G^{+++}$ and $G^{+0+}$ are high, which explains why the like-signed helical velocity field and its compressive part play an important role for this feature. However, for $k\approx q$, $G^{+0+}$ is vanishing, which explains why the NLIT is more non-local for the compressive velocity field as compared to the like-signed helical one (see figure \ref{fig:tsfamp_hm_M7c_cuts_PPP_PCP}.$(c)$, where the inverse transfer for the $\PCP$ term is visible only for shells $K \leq 30$ whereas it is visible for all shells $K \leq 50$ for the $\PPP$ one). As for $G^{+-+}$, it is vanishing in this region so that the role of the opposite-signed helical velocity field is small for this feature.
		\item Finally, all three terms have a non-vanishing geometric factor at low $p$, corresponding to the large-scale velocity field. This is why they all contribute to the LDT. The reason why the compressive velocity field plays the leading role for this feature cannot be explained alone by the fact that more energy is associated with it, as compared to the positively or negatively helical components (see figure \ref{fig:EkVPKMcomp}.$(d)$). Indeed, for the M11s run (see section \ref{sec:aboutMXIs}), the contributions of the $\PPP$ and $\PCP$ terms to the LDT have similar amplitudes, even though significantly more energy is contained in the positively helical velocity field for this run (figure \ref{fig:EkVPKMcomp}.$(b)$). The $\PCP$ term is favoured geometrically because $(i)$ on the $q=p+k$ and $q=-p+k$ lines, $G^{+0+}=2$, whereas $G^{+++}=G^{+-+}=0$ and $(ii)$ the $q=k$ horizontal line, where $G^{+++}$ and $G^{+-+}$ are high at low $p$, does not play a role for shell-to-shell transfers between shells $K$ and $Q$ since $\TsfMhl(K,K)=0$. However, as shown in section \ref{sec:results_otherruns}, even though the $\PCP$ term is favoured geometrically, it can play a less important role depending on the energy repartition among the helical velocity components.
	\end{itemize}

	The facts that the $\PPP$ transfers, mediated by the velocity field of like-signed helicity, are more efficient and more non-local than the $\PNP$ ones, where the mediating field has an opposite-signed helicity have already been predicted in the incompressible case in \cite{LSM17}.

	The analysis above assumed that the helically-decomposed \textit{shell-to-shell} transfer rates $\TsfMhlHELQPK$ have a similar shape as the \textit{single triad} geometric factors' one, which is of course not guaranteed. The magnetic helicity at a wavevector $\vk$ can be written $\mhelFk=\frac{1}{k}(|\FmagFhvkP|^2-|\FmagFhvkM|^2)$, with $\FmagFhvkPMvec=\FmagFhvkPM\vhhpm$ the helically-decomposed magnetic field, whose time evolution is governed by:

\beq
	\pat \mhelFk=\frac{1}{k}( 2 \Re(\FmagFhvkP^* \pat \FmagFhvkP + \FmagFhvkM^* \pat \FmagFhvkM)),
\eeq

	where $\Re(z)$ denotes the real part of the complex number $z$. Hence, according to relation \eqref{eq:pathelb}, the magnetic helicity transfers depend, apart from the phase information and the geometric factor, on the moduli of the interacting helical modes in the triads. In order to isolate the geometric factors' role, slices at particular $\KZ$ of the different helical components $\TsfMhlHELBASEQPKZ{\KZ}$ normalised by the typical interacting fields' moduli are considered:

\beq
	\label{eq:NormTsfMhldef}
	\TsfMhlHELBASEQPKZnormmodule{\KZ}=\frac{\TsfMhlHELBASEQPKZ{\KZ}}{2\sqrt{2\emagFhSKZ\sekinFhSP\emagFhSQ}},
\eeq

with $\emagFhSQ$ the power spectrum of the magnetic field $\hsQ$-helical part \corrT{(obtained by projecting $\FmagF$ on the $(\vhhs{\hsq}{\vq})$ helical eigenvectors in Fourier space)} and similarly for the other two energetic contributions. An average of all the $K \in [5,50]$ slices of $\TsfMhlHELBASEQPKZnormmodule{K}$ in one plot is shown in figure \ref{fig:triadSUM}. This plot has been obtained by normalising, resizing and merging the 46 slices present in this wavenumber shell range, through the procedure described in appendix \ref{app:mergeSlices}.
\def\plotvspaceval{0.5}
\def\plotvspacevalend{0}
	 
 \begin{figure}%
     \centering
	\vspace{\plotvspaceval em}
     \tsTHREEtsfamp
     \resizebox{\lwTHREEtsfamp}{!}{\input{\grpathtriad_sum_M7c10_PKM_PKMnormed_surfacenormed_absnormed_jfm.tex}}

	\vspace{\plotvspacevalend em}

     \caption[Small caption]{\footnotesize{Average of the forty six $\TsfMhlHELBASEQPKZnormmodule{\KZ}$ slices for $\KZ \in [5,50]$, obtained through the procedure described in appendix \ref{app:mergeSlices}. Each plot is normalised by its own maximum absolute value.}}
     \label{fig:triadSUM}
 \end{figure}

\def\plotvspaceval{0}
\def\plotvspacevalend{0}

	The geometric triad factors' shape is very well reflected in figure \ref{fig:triadSUM} for the $\PPP, \PNP$ and $\PCP$ terms. In this figure, the blue/yellow corner at small $P/K$ corresponds to the LDT: it is negative for $Q/K>1$ and positive for $Q/K<1$ which means that magnetic helicity is transferred from larger to smaller scales, whereas the reversal of colours at higher $P/K$ corresponds to inverse transfers (LIT and NLIT).

	This illustrates why the geometric triad factor moduli govern the role and relative importance of the dominant helical contributions with respect to shell-to-shell magnetic helicity transfers.%

\subsection{Energetic transfers}
\def\plotvspaceval{0.5}
\def\plotvspacevalend{0}
	 
 \begin{figure}%
     \centering
	\vspace{\plotvspaceval em}
     \tsTHREEtsfamp
     \resizebox{\lwTHREEtsfamp}{!}{\input{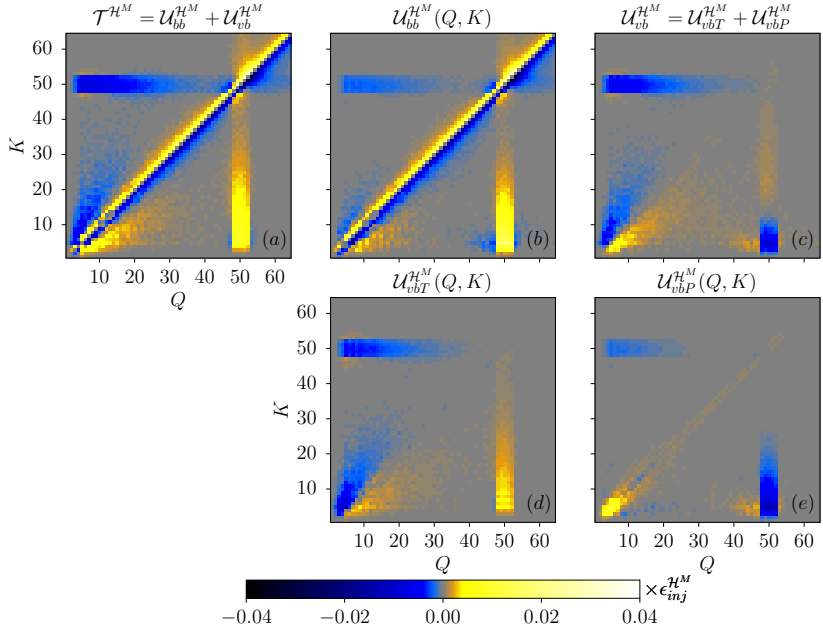}}

	\vspace{\plotvspacevalend em}

     \caption[Small caption]{\footnotesize{Decomposition of the magnetic helicity transfer rate from shell $Q$ to shell $K$ (subfigure $(a)$, which is the same as figure \ref{fig:tsfamp_hm_M7c}) in different contributions: $(b)$ associated with $\BBa$ exchanges, $(c)$ associated with $\VBa$ exchanges, the latter being the sum of: $(d)$ contributions from the magnetic tension and $(e)$ from the magnetic pressure.}}
     \label{fig:tsfamp_hm_M7c_MM_KM}
 \end{figure}

\def\plotvspaceval{0}
\def\plotvspacevalend{0}

\corr{Using the decomposition of the magnetic helicity transfer function in contributions related to energetic transfers (relations \eqref{eq:tsfmhlEneA}-\eqref{eq:tsfmhlEneB}, plotted in figure \ref{fig:tsfamp_hm_M7c_MM_KM}) allows to see that the LDT is essentially associated with a direct cascade of magnetic energy within the magnetic energy reservoir (hereafter labelled as ``$\BBa$'' exchanges). On the contrary, the LIT and the NLIT are associated with both $\BBa$ and $\VBa$ exchanges (where ``$\VBa$'' represents transfers between the magnetic and kinetic energy reservoirs). As expected from the analysis in section \ref{sec:results_helical}, the LIT is mostly associated with magnetic tension, since the magnetic pressure is exclusively associated with the compressive part of the velocity field.}

	\corr{A more detailed analysis of the roles of $\BBa$ and $\VBa$ energetic exchanges, decomposed in velocity-field helical contributions, is given in appendix \ref{app:enetsf}. Its conclusion is that the like-signed velocity field helical part not only mediates the LIT and NLIT but also leads to $\VBa$ exchanges in favour of the inverse transfer, whereas the $\VBa$ exchanges with the compressive part of the velocity field play against the inverse transfer. The overall participation of the velocity field compressive part to the inverse transfer remains however positive because its mediating role in $\BBa$ exchanges overrules its role in the counteracting $\VBa$ exchanges.}

\subsection{Other runs}
\label{sec:results_otherruns}
\newcommand{\vspacevalendThreeCut}{0}
\newcommand{\vspacevalThreeCut}{0}

\def\plotvspacevalend{\vspacevalendThreeCut}
\def\plotvspaceval{\vspacevalThreeCut}
	 
 \begin{figure}%
\begin{center}
	M11s
\end{center}
{
     \centering
	\vspace{\plotvspaceval em}
{
     \huge
     \resizebox{1.\linewidth}{!}{\input{\grpathtsfamp_M12s_2D_TOT_PPP_PCP_cutK30.tex}}	
}
	\vspace{\plotvspacevalend em}
{
     \caption[Small caption]{\footnotesize{$(a \sm c)$ Magnetic helicity transfer rates between the shells $Q$ and $K$ and their greatest helical contributions for the M11s run at an instant in time when $\spectransMainInstant$. $(d)$ Cuts of subfigures $(a \sm c)$ at $\KZ=\cutB$. The transfers are expressed in units of the magnetic helicity injection rate $\mhelinj$.}}
     \label{fig:tsfamp_M12s_2D_TOT_PPP_PCP_cutK30}
}
}
 \end{figure}

\def\plotvspacevalend{0}
\def\plotvspaceval{0}

\def\plotvspaceval{0.5}
	 
 \begin{figure}%
\begin{center}
	M1c
\end{center}
{
     \centering
	\vspace{\plotvspaceval em}
{
     \Huge
     \resizebox{1.\linewidth}{!}{\input{\grpathtsfamp_hm_M1c_PPP_PCP_cuts.tex}}	
}
	\vspace{\plotvspacevalend em}
{
     \caption[Small caption]{\footnotesize{$(a \sm c)$ Magnetic helicity transfer rates between the shells $Q$ and $K$ and their greatest helical contributions for the M1c run at an instant when $\spectransMainInstant$. Please note that the extremes of the colour bar of subfigure $(b)$ are an order of magnitude lower as compared to subfigures $(a)$ and $(c)$. $(d \sm f)$ Cuts of the subfigures $(a \sm c)$ at $\KZ=\cutA$, $\KZ=\cutB$ and $\KZ=\cutC$ respectively. The transfers are expressed in units of the magnetic helicity injection rate $\mhelinj$.}}
     \label{fig:tsfamp_hm_M1c_PPP_PCP_cuts}
}
}
 \end{figure}

\def\plotvspaceval{0}

\def\plotvspacevalend{\vspacevalendThreeCut}
\def\plotvspaceval{\vspacevalThreeCut}
	 
 \begin{figure}%
\begin{center}
	M01s4
\end{center}
{
     \centering
	\vspace{\plotvspaceval em}
{
     \huge
     \resizebox{1.\linewidth}{!}{\input{\grpathtsfamp_M01s4_PPP_PNP.tex}}	
}
	\vspace{\plotvspacevalend em}
{
     \caption[Small caption]{\footnotesize{$(a \sm c)$ Magnetic helicity transfer rates from shell $Q$ to shell $K$ for the M01s4 run at an instant when $\spectransMainInstant$, as well as the biggest helical contributions. $(d)$ Cuts of subfigures $(a \sm c)$ at $\KZ=30$. The transfers are expressed in units of the magnetic helicity injection rate $\mhelinj$.}}
     \label{fig:tsfamp_M01s4_PPP_PNP}
}
}
 \end{figure}

\def\plotvspacevalend{0}
\def\plotvspaceval{0}

The other runs, M01s4, M11s and M1c, are also considered at an instant when $\IscHm\approx \Lbox/10$. Even though the compressibility of these runs extends over a wide range, they exhibit a lot of common properties with the M8c run: the same three features (LDT, LIT and NLIT) are present and mediated by the velocity field at the same scales. The LDT is always mostly associated with $\BBa$ exchanges whereas the LIT and NLIT are associated with both $\BBa$ and $\VBa$ exchanges. The differences observed are essentially quantitative and concern the different helical contributions' importance. They can be explained through the energy repartition in the velocity field's helical components, given in figure \ref{fig:EkVPKMcomp}.

	\label{sec:aboutMXIs}For the M11s run (figure \ref{fig:tsfamp_M12s_2D_TOT_PPP_PCP_cutK30}), both the velocity field's compressive and like-signed helical parts contribute in equal proportions to the LDT, contrary to the situation for the M8c run. For this run indeed, the positively helical velocity field entails significantly more energy $\sekinFhP$ than its compressive part $\sekinFhC$. This confirms that the compressive velocity field is indeed geometrically favoured for the LDT.

	The other extreme happens for the M1c run, for which the $\sekinFhC/\sekinFhP$ ratio is very large at large scales and small at small scales. As a consequence, the LDT has a very large magnitude as compared to the NLIT and the LIT (figures \ref{fig:tsfamp_hm_M1c_PPP_PCP_cuts}.$(a\sm c)$, where the colour bar extremes in the $(b)$ plot are an order of magnitude smaller) and the $\PCP$ term does not contribute much to the NLIT (subfigures $(c,f)$).

	Lastly, the LDT is carried by both velocity field's helical components for the subsonic M01s4 run (figure \ref{fig:tsfamp_M01s4_PPP_PNP}), since there is only a negligible amount of energy in the compressive velocity field. For this run, both $\PPP$ and $\PNP$ terms have the same importance with respect to the LDT because similar amounts of energy are present in the positively and negatively helical parts of the velocity field and the geometric factors' magnitude $G^{+++}$ and $G^{+-+}$ (see relation \eqref{eq:Gincomp}) have very close values for $p \ll k \approx q$.

\section{Conclusion}
\label{sec:conclusion}
	The helically-decomposed shell-to-shell analysis reveals the presence of three distinct phenomena occuring in the global picture of the inverse transfer of magnetic helicity and sheds some light on their origin. Since one sign of magnetic helicity dominates the system at all scales, the role of the different helical components of the velocity field as well as that of its compressive part can be distinguished. The analysis has been performed on particular states taken from direct numerical simulations of magnetic helicity inverse transfer in compressible isothermal MHD flows. The range of Mach numbers varies from subsonic to RMS Mach numbers of the order of 10, with either a purely solenoidal or compressive mechanical large-scale forcing. The results are consistent with previous research done in the incompressible case regarding the transfer's direction and locality \cite{AMP06} and relative strength and locality of triad interactions involving the like-signed and opposite-signed helical part of the velocity field \cite{LSM17}.

	For all the runs considered, a local inverse transfer (LIT), a non-local inverse transfer (NLIT) and a local direct transfer (LDT) are observed: %

\begin{enumerate}
	\item The LIT is caused essentially by magnetic stretching involving the solenoidal velocity field at intermediate scales. The like-signed helical velocity field plays here the dominant role. It is associated with both magnetic$\lra$magnetic ($\BBa$) and kinetic$\lra$magnetic ($\VBa$) energetic exchanges.

	\item The leading role in the NLIT is taken by the small-scale velocity field, through both the like-signed helical velocity field and its compressive part. Their roles are however very different: while the like-signed helical velocity field participates in the inverse transfer by $\VBa$ exchanges at larger scales and by mediating $\BBa$ transfers, the compressive part of the velocity field acts energetically against the inverse transfer, but plays an important mediating role in the $\BBa$ transfers, leaving a net positive contribution to the NLIT.

	\item Finally, the LDT is caused by the large-scale velocity field. It is essentially associated with a direct cascade of magnetic energy. The compressive part of the velocity field is geometrically favoured for the LDT even though both velocity helical components, which have geometrically a similar importance, can take the leading role depending on the specific kinetic energy repartition among the helical and compressive modes.
\end{enumerate}

	\corrT{Thus the velocity field's compressive part, which is not present in the incompressible case, plays an important role in the LDT and the NLIT, which affects the magnetic helicity scaling properties \cite{TEM20}.}

	The locality and strength of these phenomena can surprisingly well be explained by the geometric triad helical factors \cite{WAL92}, which have been extended for compressible MHD. 
	Since magnetic helicity is not sign-definite, in astrophysical systems of interest, magnetic helical components of mixed signs are expected to be present. In this context, the heterochiral terms (labelled ``$\SPS, \SNS$'' and ``$\SCS$'') may play a greater role, even though they are more difficult to interpret. Nevertheless, the study presented here in the simpler case of one helical sign injection provides insights which should help the interpretation of future results in more general cases.

The authors acknowledge the North-German Supercomputing Alliance (HLRN) for providing HPC resources that have contributed to the research results reported in this paper. Computing resources from the Max Planck Computing and Data Facility (MPCDF) are also acknowledged. JMT gratefully acknowledges support by the Berlin International Graduate School in Model and Simulation based Research (BIMoS).

Declaration of Interests. The authors report no conflict of interest.

\appendix

\section{Geometric factor derivation}
\label{app:Qgeom}

\def\vspacetikzfig{-3}
        
 \begin{figure}%
     \centering
     \large
 \resizebox{0.75\linewidth}{!}{\input{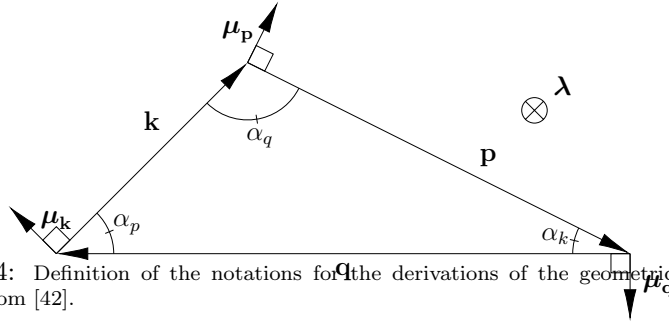}}

\vspace{\vspacetikzfig em}

     \caption[Small caption]{\footnotesize{Definition of the notations for the derivations of the geometric triad factor. Remade from \cite{WAL92}.}}
     \label{fig:sketch_triad}
 \end{figure}

\def\vspacetikzfig{0}

	The geometric factor appearing in relation \eqref{eq:pathelb} has been derived in \cite{WAL92} in the incompressible case. The derivations are reviewed here, keeping the same notations shown in figure \ref{fig:sketch_triad}. The helical eigenvectors $\vhhs{\hsk}{\vk}, \vhhs{\hsp}{\vp} $ and $\vhhs{\hsq}{\vq}$ are expressed for each triad in a well-chosen basis:

\newcommand{\ha}{\phi}

\beq
	\vhhs{\hsm}{\vm}=e^{i\hsm\ha_m}(\vlambda+i\hsm\vmu_{\vm}),
\eeq

	with $\vm \in \{\vk,\vp,\vq\}$, $\ha_m$ a certain angle and $\vlambda$ and $\vmu_{\vm}$ the unitary vectors:

\beqa
                \vlambda&=&\frac{\vk \times \vp}{|\vk \times \vp|}=\frac{\vp \times \vq}{|\vp \times \vq|}=\frac{\vq \times \vk}{|\vq \times \vk|},\\
                \vmu_{\vneutrindex}&=&\frac{\vneutrindex \times \vlambda}{|\vneutrindex|}.
\eeqa

	In the incompressible case ($\hsk,\hsp,\hsq \in \{+,-\}$), this gives after some algebra the geometric triad factor \cite{WAL92}:

\beqa
                \!\!\!\!\!\!\!\!\triadgeomkpq&=&(\vhhs{\hsp*}{\vp} \times \vhhs{\hsq*}{\vq}) \cdot \vhhs{\hsk*}{\vk},\\
                &=&-e^{-i (\hsk \ha_k + \hsp \ha_p + \hsq \ha_q)}\hsk\hsp\hsq(\hsk \sin(\alpha_k)+\hsp \sin(\alpha_p)+\hsq \sin(\alpha_q)),\\
		&=&e^{-i \ha_s}\frac{\hsk\hsp\hsq(\hsk k+\hsp p+\hsq q)\sqrt{2k^2p^2+2p^2q^2+2q^2k^2-k^4-p^4-q^4}}{2kpq},
\eeqa
        with $\ha_s=(\hsk \ha_k + \hsp \ha_p + \hsq \ha_q)$. The last equality is obtained by using the law of sines and Heron's formula.

	In the compressible case, when $\hsp=0, \hsk=S \in \{+,-\}, \hsq=\pm S$, the geometric factor becomes:
                
\beqa
	\triadgeomkpqC&=&(\frac{\vp}{p} \times \vhhs{\hsq*}{\vq}) \cdot \vhhs{\hsk*}{\vk},\\
	&=&Sie^{-Si(\ha_k\pm\ha_q)}(\cos(\alpha_q)\mp \cos(\alpha_k)),\\
	&=&Sie^{-Si(\ha_k\pm\ha_q)}\frac{(q\mp k)(p^2-k^2-q^2\mp 2qk)}{2kpq},
\eeqa

	where the last equality is obtained using the cosine rule $\cos(\alpha_k)=\frac{kp^2+kq^2-k^3}{2kpq}$.

\section{Technical details: merging transfer functions' slices}

\label{app:mergeSlices}

 \begin{figure}%
     \centering
     \large
 \resizebox{1.\linewidth}{!}{\input{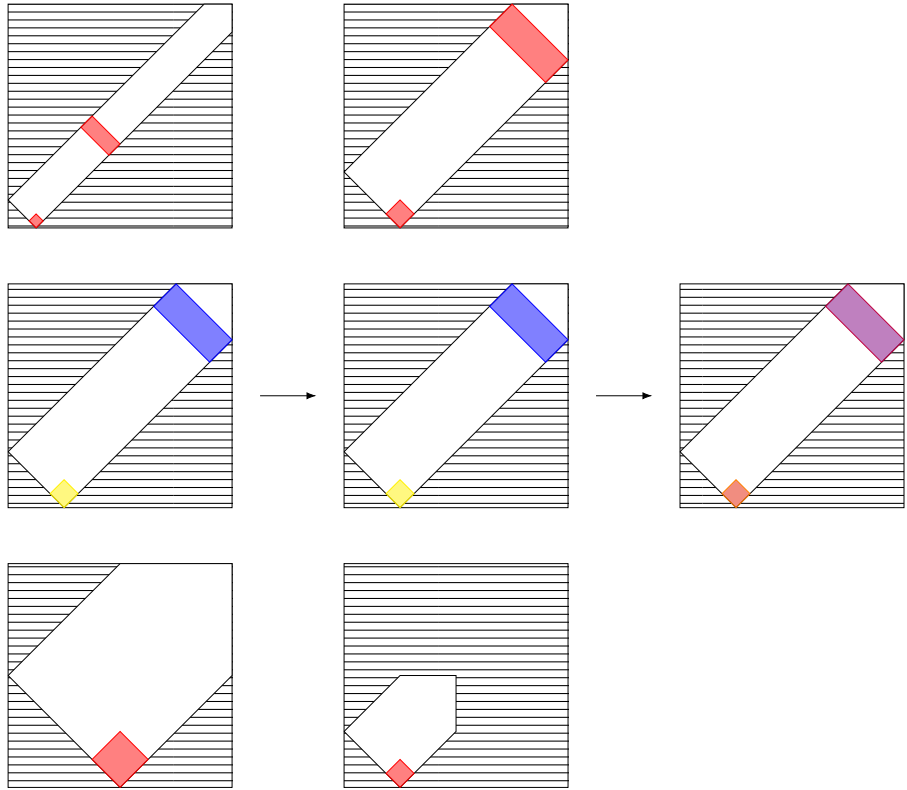}}

\vspace{\vspacetikzfig em}

     \caption[Small caption]{\footnotesize{Illustration of the algorithm generating figure \ref{fig:triadSUM}. On the left: three slices being merged. In the middle, after resizing so that their angles coincide. On the right: the final merged picture. The sum of the red and blue bands from the top and middle pictures gives the purple band, which is normalised by a factor 2, since only two of the three slices contributed to it. The smaller orange square is normalised by a factor 3 since it comes from all the three slices.}}
     \label{fig:sketch_tsfamptriad}
 \end{figure}

        Figure \ref{fig:triadSUM} is obtained by merging the $\TsfMhlHELBASEQPKZnormmodule{K}$ slices (relation \eqref{eq:NormTsfMhldef}) for $K \in [5,50]$ through the following procedure (see figure \ref{fig:sketch_tsfamptriad}):

\begin{enumerate}
        \item Each slice is considered as a ``picture'' and is resized so that the angles $P=K$ and $Q=K$ arrive at the same location for each of them. The slices are normalised by the maximum of their respective absolute value so that they all have the same weight in the final merged picture. Only the slices starting $K=5$ are considered since low $K$ slices are more imprecisely resized, causing square-like artifacts.
        \item For each $(P,Q)$ coordinate, the sum of all resized pictures is considered, leading to their superposition, the new ``picture'' $S$.
        \item Each $(P,Q)$ coordinate in the $S$ picture is normalised by the amount of pictures that contributed to it (it differs for each point since all resized pictures occupy a different domain, see figure \ref{fig:sketch_tsfamptriad}).
	\item The obtained $S$ is finally normalised by its maximum absolute value.
\end{enumerate}

\section{Helically-decomposed energetic exchanges}
\label{app:enetsf}
\newcommand{\TsfAMPpA}{\sum_X \TsfEmBBPK_{\FvelhX}}
\newcommand{\TsfAMPpB}{\sum_X \TsfEmKBTPK_{\FvelhX}}
\newcommand{\TsfAMPpC}{\sum_X (\TsfEmBBvhX+\TsfEmKBTvhX)+\TsfEmKBP}

\newcommand{\TsfAMPpD}{\TsfEmBBPK_{\FvelhP}}
\newcommand{\TsfAMPpE}{\TsfEmKBTPK_{\FvelhP}}
\newcommand{\TsfAMPpF}{\sum_X \TsfEmKBTPK_{\FvelhX}+\TsfEmKBP}

\newcommand{\TsfAMPpG}{\TsfEmBBPK_{\FvelhC}}
\newcommand{\TsfAMPpH}{\TsfEmKBTPK_{\FvelhC}}
\newcommand{\TsfAMPpI}{\TsfEmKBPPK=\TsfEmKBPvhC}

\newcommand{\TsfAMPpJ}{\TsfEmBBPK_{\FvelhN}}
\newcommand{\TsfAMPpK}{\TsfEmKBTPK_{\FvelhN}}

 \begin{figure}%
     \centering
\begin{minipage}{0.8\linewidth}
     \small
 	\resizebox{\textwidth}{!}{\input{\grpathsketch_tsfEm_PPP.tex}}
	\resizebox{\textwidth}{!}{\input{\grpathsketch_tsfEm_PCP.tex}}
\end{minipage}
     \caption[Small caption]{\footnotesize{Sketch of the dominant $\VBa$ and $\BBa$ energetic exchanges, done from the plots of figure \ref{fig:tsfamp_em_M7c_KBT_KBP}, involving either $(a)$ the positively helical velocity field or $(b)$ its compressive part. Black arrows between the $K$-and $P$-axis correspond to magnetic stretching, red to magnetic pressure. The dashed lines correspond to the velocity scales mediating $\BBa$ exchanges. The thickness of the line indicates the transfer's importance. The ticks along the $P$-and $K$-axis delimit ``large'' ($K \lessapprox 10\approx \Lbox/\IscHm$, $P \lessapprox 3$), ``small'' ($K,P \gtrapprox 40$) and ``intermediate'' (in between) scales. The LIT, associated with the positively helical velocity field and with both $\VBa$ and $\BBa$ exchanges also occurs at small $K$ but is not drawn in order not to overload the sketch. For the same reason, the LDT, mediated by the large-scale velocity field, is not drawn.}}
     \label{fig:sketch_tsfEm}
 \end{figure}

 \begin{figure}%
     \centering
	\vspace{\plotvspaceval em}
     \tsTHREEtsfamp
     \resizebox{\lwTHREEtsfamp}{!}{\input{\grpathtsfamp_em_M7c10_KBTKBPBB_PK_jfm.tex}}

	\vspace{\plotvspacevalend em}

     \caption[Small caption]{\footnotesize{Several functions showing energy exchanges between magnetic and velocity fields as well as the mediating role of the velocity field, \MVIIc\ run at an instant when $\spectransMainInstant$. The plots are in units of the magnetic energy injection rate $\EMinj$ (not to be confused with the magnetic helicity injection rate $\mhelinj$).}}
     \label{fig:tsfamp_em_M7c_KBT_KBP}
 \end{figure}

	This appendix aims at analysing in more detail the role of the different helical components of the velocity field, as well as its compressive part, separating their role with respect to the mediation of purely magnetic$\lra$magnetic ($\BBa$) energy exchanges and exchanges between the kinetic and magnetic energy reservoirs ($\VBa$ exchanges), splitting the latter in contributions from magnetic stretching and magnetic pressure.
	For this purpose, the energy transfer functions given in relations \eqref{eq:tsfemvb} and \eqref{eq:magpresstretch} can be rewritten, when shell-filtering all interacting fields and considering a certain velocity helical component $\FvelhX$ with $X\in\{P,N,C\}$ corresponding respectively to positively helical, negatively helical and compressive parts:

\beqa
	\label{eq:tsfembb3fil}
	\TsfEmBBQPKcomp_{\FvelhX}&=& \volavg{-\FmagK \cdot (\FvelhXP \cdot \nabla) \FmagQ - \frac{1}{2} (\FmagK \cdot \FmagQ) \nabla \cdot \FvelhXP},\\
	\TsfEmKBTQPK_{\FvelhX} &=& \volavg{\FmagK \cdot (\FmagQ \cdot \nabla) \FvelhXP},\\
	\TsfEmKBPQPK_{\FvelhX} &=& \volavg{-\frac{1}{2} (\FmagK \cdot \FmagQ) \nabla \cdot \FvelhXP}.
\eeqa

	Summing these functions of three variables $(Q,P,K)$ appropriately allows to investigate particular aspects of the dynamics. For example, a sum over $P$ and the helical component $X$ of relation \eqref{eq:tsfembb3fil} gives the $\BBa$ shell-to-shell transfer rates between shells $Q$ and $K$ (relation \eqref{eq:tsfembb}). In order to assess the importance of $\FvelhXP$, a sum over the $Q$ index is hence needed. The sums $\sum_Q \TsfEmKBTQPK_{\FvelhX}=\TsfEmKBTPK_{\FvelhX}$ and $\sum_Q \TsfEmKBPQPK_{\FvelhX}=\TsfEmKBPPK_{\FvelhX}$ correspond to conversion of kinetic energy associated with the $\FvelhX$ field at shell $P$ to magnetic energy at shell $K$, through magnetic stretching and magnetic pressure respectively. On the other hand, $\sum_Q \TsfEmBBQPK_{\FvelhX}=\TsfEmBBPK_{\FvelhX}$ corresponds to the mediating role of the $\FvelhX$ field at shell $P$ in $\BBa$ transfers to shell $K$, similarly to $\TsfMhlPK$ for magnetic helicity transfers (see relation \ref{eq:MhlPK}).

	The dominant phenomena are summarised as a sketch, figure \ref{fig:sketch_tsfEm}, which has been drawn from the plots in figure \ref{fig:tsfamp_em_M7c_KBT_KBP}. The analysis of relevant transfer functions show that the electromotive forcing, which injects positively helical magnetic fluctuations, leads through magnetic stretching (figure \ref{fig:tsfamp_em_M7c_KBT_KBP}.$(e)$) to a strong $\FvelhP$ at small scales, which plays in turn an important role in the inverse transfer by mediating the NLIT (subfigure $(d)$) but as well by being converted into magnetic energy at intermediate and large scales (subfigure $(e)$). At intermediate scales, the $\FvelhP$ field contributes to the LIT through energetic exchanges and mediation of $\BBa$ transfers. Regarding the compressive part of the velocity field $\FvelhC$, the $\VBa$ energetic exchanges play globally against the inverse transfer: even though magnetic stretching transforms kinetic energy to magnetic energy at larger scales (subfigure $(h)$), the transfers due to magnetic pressure are in the other direction and are stronger (subfigure $(i)$). The overall participation of the small-scale $\FvelhC$ in the inverse transfer is however positive, due to its very important mediating role in $\BBa$ exchanges (subfigure $(g)$). As for the $\FvelhN$ field, even though its role in $\VBa$ energetic exchanges is relatively small, its mediating role at intermediate and large scales looks similar to that of the $\FvelhP$ field.

\bibliography{biblio}

\begin{thebibliography}{10}

\bibitem{ALE17}
A.~Alexakis.
\newblock Helically decomposed turbulence.
\newblock {\em Journal of Fluid Mechanics 812}, pages 752--770, 2017.

\bibitem{ALB18}
A.~Alexakis and L.~Biferale.
\newblock Cascades and transitions in turbulent flows.
\newblock {\em Physics Reports 767-769}, pages 1--101, 2018.

\bibitem{AMP05}
A.~Alexakis, P.~D. Mininni, and A.~Pouquet.
\newblock Shell-to-shell energy transfer in magnetohydrodynamics. {I}. {S}teady
  state turbulence.
\newblock {\em Physical Review E 72, 046301}, 2005.

\bibitem{AMP06}
A.~Alexakis, P.~D. Mininni, and A.~Pouquet.
\newblock On the inverse cascade of magnetic helicity.
\newblock {\em The Astrophysical Journal 640}, pages 335--343, 2006.

\bibitem{ALF42}
H.~Alfv\'{e}n.
\newblock On the existence of electromagnetic-hydrodynamic waves.
\newblock {\em Arkiv f\"{o}r Matematik, Astronomi och Fysik 29B(2)}, pages
  1--7, 1942.

\bibitem{AEY10}
H.~Aluie and G.~L. Eyink.
\newblock Scale locality of magnetohydrodynamic turbulence.
\newblock {\em Physical Review Letters 104, 081101}, 2010.

\bibitem{BAP99}
D.~Balsara and A.~Pouquet.
\newblock The formation of large-scale structures in supersonic
  magnetohydrodynamic flows.
\newblock {\em Physics of Plasmas Vol. 6 No. 1}, pages 89--99, 1999.

\bibitem{BAL10}
D.~S. Balsara.
\newblock Multidimensional {HLLE} {R}iemann solver: Application to {Euler} and
  magnetohydrodynamic flows.
\newblock {\em Journal of Computational Physics 229}, pages 1970--1993, 2010.

\bibitem{BER99}
M.~A. Berger.
\newblock Introduction to magnetic helicity.
\newblock {\em Plasma Physics and Controlled Fusion 41}, pages B167--B175,
  1999.

\bibitem{BEM87}
J.~W. Bieber, P.~A. Evenson, and W.~H. Matthaeus.
\newblock Magnetic helicity of the {P}arker field.
\newblock {\em The Astrophysical Journal, 315}, pages 700--705, 1987.

\bibitem{BRA01}
A.~Brandenburg.
\newblock The inverse cascade and nonlinear alpha-effect in simulations of
  isotropic helical hydromagnetic turbulence.
\newblock {\em The Astrophysical Journal 550}, pages 824--840, 2001.

\bibitem{BRL13}
A.~Brandenburg and A.~Lazarian.
\newblock Astrophysical hydromagnetic turbulence.
\newblock {\em Space Science Reviews 178}, pages 163--200, 2013.

\bibitem{BRS05}
A.~Brandenburg and K.~Subramanian.
\newblock Astrophysical magnetic fields and nonlinear dynamo theory.
\newblock {\em Physics Reports 417}, pages 1--209, 2005.

\bibitem{BUH14}
P.~Buchm\"{u}ller and C.~Helzel.
\newblock Improved accuracy of high-order {WENO} finite volume methods on
  cartesian grids.
\newblock {\em Journal of Scientific Computing 61}, pages 343--368, 2014.

\bibitem{CHB01}
M.~Christensson and M.~Hindmarsh.
\newblock Inverse cascade in decaying three-dimensional magnetohydrodynamic
  turbulence.
\newblock {\em Physical Review E 64, 056405}, 2001.

\bibitem{ELS04}
B.~G. Elmegreen and J.~Scalo.
\newblock Interstellar turbulence {I}: Observations.
\newblock {\em Annual Review of Astronomy and Astrophysics 42}, pages 211--273,
  2004.

\bibitem{EMO00}
D.~F. Escande, P.~Martin, S.~Ortolani, A.~Buffa, P.~Franz, L.~Marrelli,
  E.~Martines, G.~Spizzo, S.~Cappello, A.~Murari, R.~Pasqualotto, and P.~Zanca.
\newblock Quasi-single-helicity reversed-field-pinch plasmas.
\newblock {\em Physical Review Letters 85, No. 8}, pages 1662--1665, 2000.

\bibitem{EVH88}
C.~R. Evans and J.~F. Hawley.
\newblock Simulation of magnetohydrodynamic flows: a constrained transport
  method.
\newblock {\em The Astrophysical Journal 332}, pages 659--677, 1988.

\bibitem{FED13}
C.~Federrath.
\newblock On the universality of supersonic turbulence.
\newblock {\em Monthly Notices of the Royal Astronomical Society 436}, pages
  1245--1257, 2013.

\bibitem{FRD10}
C.~Federrath, J.~Roman-Duval, R.~S. Klessen, W.~Schmidt, and M.-M. Mac~Low.
\newblock Comparing the statistics of interstellar turbulence in simulations
  and observations, solenoidal versus compressive turbulence forcing.
\newblock {\em Astronomy and Astrophysics 512, A81}, 2010.

\bibitem{FPL75}
U.~Frisch, A.~Pouquet, J.~L\'{e}orat, and A.~Mazure.
\newblock Possibility of an inverse cascade of magnetic helicity in
  magnetohydrodynamic turbulence.
\newblock {\em Journal of Fluid Mechanics 68 Part 4}, pages 769--778, 1975.

\bibitem{GOS17}
P.~Grete, B.~W. O'Shea, K.~Beckwith, W.~Schmidt, and A.~Christlieb.
\newblock Energy transfer in compressible magnetohydrodynamic turbulence.
\newblock {\em Physics of Plasmas 24, 092311}, 2017.

\bibitem{KET08}
D.~I. Ketcheson.
\newblock Highly efficient strong stability-preserving {Runge-Kutta} methods
  with low-storage implementations.
\newblock {\em Society for Industrial and Applied Mathematics Journal on
  Scientific Computing 30, No. 4}, pages 2113--2136, 2008.

\bibitem{KWN13}
A.~G. Kritsuk, R.~Wagner, and M.~L. Norman.
\newblock Energy cascade and scaling in supersonic isothermal turbulence.
\newblock {\em Journal of Fluid Mechanics Vol. 729, R1}, 2013.

\bibitem{KUR96}
A.~Kumar and D.~M. Rust.
\newblock Interplanetary magnetic clouds, helicity conservation, and
  current-core flux-ropes.
\newblock {\em Journal of Geophysical Research 101}, pages 667--684, 1996.

\bibitem{LPC09}
T.~Lessinnes, F.~Plunian, and D.~Carati.
\newblock Helical shell models for {MHD}.
\newblock {\em Theoretical and Computational Fluid Dynamics 23}, pages
  439--450, 2009.

\bibitem{LPR99}
D.~Levy, G.~Puppo, and G.~Russo.
\newblock Central {WENO} schemes for hyperbolic systems of conservation laws.
\newblock {\em Mathematical Modelling and Numerical Analysis 33, No. 3}, pages
  547--571, 1999.

\bibitem{LBM16}
M.~Linkmann, A.~Berera, M.~McKay, and J.~J\"{a}ger.
\newblock Helical mode interactions and spectral transfer processes in
  magnetohydrodynamic turbulence.
\newblock {\em Journal of Fluid Mechanics 791}, pages 61--96, 2016.

\bibitem{LSM17}
M.~Linkmann, G.~Sahoo, M.~McKay, A.~Berera, and L.~Biferale.
\newblock Effects of magnetic and kinetic helicities on the growth of magnetic
  fields in laminar and turbulent flows by helical {F}ourier decomposition.
\newblock {\em The Astrophysical Journal 836:26}, 2017.

\bibitem{LOW94}
B.~C. Low.
\newblock Magnetohydrodynamic processes in the solar corona: Flares, coronal
  mass ejections, and magnetic helicity.
\newblock {\em Physics of Plasmas 1}, pages 1684--1690, 1994.

\bibitem{MAL09}
S.~K. Malapaka.
\newblock {\em A Study of Magnetic Helicity in Decaying and Forced {3D-MHD}
  Turbulence}.
\newblock PhD thesis, Universit\"{a}t Bayreuth, 2009.

\bibitem{COC11}
P.~McCorquodale and P.~Colella.
\newblock A high-order finite-volume method for conservation laws on locally
  refined grids.
\newblock {\em Communications in Applied Mathematics and Computational Science
  6, No. 1}, pages 1--25, 2011.

\bibitem{MFP81}
M.~Meneguzzi, U.~Frisch, and A.~Pouquet.
\newblock Helical and nonhelical turbulent dynamos.
\newblock {\em Physical Review Letters 47, No. 15}, pages 1060--1064, 1981.

\bibitem{MMB12}
W.-C. M\"{u}ller, S.~K. Malapaka, and A.~Busse.
\newblock Inverse cascade of magnetic helicity in magnetohydrodynamic
  turbulence.
\newblock {\em Physical Review E 85, 015302}, 2012.

\bibitem{PFL76}
A.~Pouquet, U.~Frisch, and J.~L\'{e}orat.
\newblock Strong {MHD} helical turbulence and the nonlinear dynamo effect.
\newblock {\em Journal of Fluid Mechanics 77, part 2}, pages 321--354, 1976.

\bibitem{POP78}
A.~Pouquet and G.~S. Patterson.
\newblock Numerical simulation of helical magnetohydrodynamic turbulence.
\newblock {\em Journal of Fluid Mechanics 85, part 2}, pages 305--323, 1978.

\bibitem{RUS61}
V.~V. Rusanov.
\newblock The calculation of the interaction of non-stationary shock waves with
  barriers.
\newblock {\em Zhurnal Vychislitel'noi Matematiki i Matematicheskoi Fiziki,
  1:2}, pages 267--279, 1961.
\newblock English: USSR Computational Mathematics and Mathematical Physics,
  1:2, pp. 304-320, 1962.

\bibitem{TEI20}
J.-M. Teissier.
\newblock {\em Magnetic helicity inverse transfer in isothermal supersonic
  magnetohydrodynamic turbulence}.
\newblock PhD thesis, Technische Universit\"{a}t Berlin, 2020.

\bibitem{TEM20}
J.-M. Teissier and W.-C. M\"{u}ller.
\newblock Inverse transfer of magnetic helicity in supersonic
  magnetohydrodynamic turbulence.
\newblock {\em Journal of Physics: Conference Series 1623, 012011}, 2020.

\bibitem{VTH19}
P.~S. Verma, J.-M. Teissier, O.~Henze, and W.-C. M\"{u}ller.
\newblock Fourth-order accurate finite-volume {CWENO} scheme for astrophysical
  {MHD} problems.
\newblock {\em Monthly Notices of the Royal Astronomical Society 482}, pages
  416--437, 2019.

\bibitem{VIC01}
E.~T. Vishniac and J.~Cho.
\newblock Magnetic helicity conservation and astrophysical dynamos.
\newblock {\em The Astrophysical Journal 550}, pages 752--760, 2001.

\bibitem{WAL92}
F.~Waleffe.
\newblock The nature of triad interactions in homogeneous turbulence.
\newblock {\em Physics of Fluids A: Fluid Dynamics 4}, pages 350--363, 1992.

\end{thebibliography}
\bibliographystyle{plain}

\end{document}